\documentclass[prd,preprintnumbers,amsmath,amssymb,nofootinbib]  {revtex4}
\usepackage[dvips,final]{graphicx}
\usepackage{amsmath}
\usepackage{amssymb}
\usepackage{url}
\usepackage{subfigure}
\usepackage{textcomp}
\usepackage{graphicx}
\usepackage{epsfig}
\usepackage{bm}
\usepackage{dcolumn}
\usepackage[usenames]{color}
\usepackage{epstopdf}

\usepackage{graphicx}
\newcommand{\beq}{\begin{equation}}
\newcommand{\eeq}{\end{equation}}
\newcommand{\bea}{\begin{eqnarray}}
\newcommand{\eea}{\end{eqnarray}}
\newcommand{\bi}{\begin{itemize}}
\newcommand{\ei}{\end{itemize}}

\newcommand{\eend}{\textrm{end}}
\newcommand{\re}{\textrm{re}}
\newcommand{\s}{\textrm{s}}
\newcommand{\intt}{\textrm{int}}

\def\beq{\begin{equation}}
\def\eeq{\end{equation}}
\def\bea{\begin{eqnarray}}
\def\eea{\end{eqnarray}}

\makeatletter
\newcommand*{\rom}[1]{\expandafter\@slowromancap\romannumeral #1@}
\makeatother

\begin{document}

\title{ CMB and Reheating Constraints to  $\alpha\,$-attractor Inflationary Models }

\author{Mehdi Eshaghi$^{1,2}$}
\email{m.eshaghi(AT)sci.ui.ac.ir}

\author{Moslem Zarei$^{3,5}$}
\email{m.zarei(AT)cc.iut.ac.ir}

\author{Nematollah Riazi$^{4}$}
\email{n_riazi(AT)sbu.ac.ir}

\author{ Ahmad Kiasatpour $^{1}$}
\email{akiasat(AT)sci.ui.ac.ir}

\affiliation{ $^1$ Department of Physics , Faculty of Science, University of Isfahan, Isfahan, 81746-73441, Iran}

\affiliation{ $^2$ Astrophysics Sector, SISSA, Via Bonomea 265, I-34136 Trieste, Italy}

\affiliation{ $^3$ Department of Physics, Isfahan University of Technology, Isfahan 84156-83111, Iran}

\affiliation{$^4$ Department of Physics, Shahid Beheshti University, Tehran 19839, Iran}

\affiliation{$^5$ School of Astronomy, Institute for Research in Fundamental Sciences (IPM),P.~O.~Box 19395-5531,Tehran, Iran}


\date{\today}

 \begin{abstract}
After Planck 2013, a broad class of inflationary models called $\alpha\,$-attractors was developed which has universal observational predictions. For small values of the parameter $\alpha$, the models have good consistency with the recent CMB data. In this work, we first calculate analytically (and verify numerically) the predictions of these models for spectral index, $n_{\textrm{s}}$, and tenso-to-scalar ratio, $r$, and then using BICEP2/Keck 2015 and Planck 2015 data we impose constraints on $\alpha\,$-attractors. Then, we study the reheating in $\alpha$-attractors. The reheating temperature, $T_{\textrm{re}}$, and the number of e-folds during reheating, $N_{\textrm{re}}$,
 are calculated as functions of $n_{\textrm{s}}$. Using these results, we determine the range of free parameter of two classes of $\alpha\,$-attractors which satisfy the constraints of recent CMB data.

 \end{abstract}

\maketitle

\section{Introduction}

The high-precision Cosmic Microwave Background (CMB) data released by the Planck collaboration \cite{Ade:2015lrj} has more tightly constrained inflationary models. Based on these results,  single field inflationary models with plateau-like shapes are more favored by Planck+WMAP data \cite{Ade:2015lrj,Ijjas:2013vea}. In general, the Planck 2015 data show that CMB temperature anisotripies and polarization can be explained by primordial Gaussian fluctuations with a nearly scale invariant power spectrum parameterized as
$\mathcal{P}(k)=A_{\textrm{s}}\left(k/k_{\ast}\right)^{n_{\textrm{s}}-1}$ where $k_{\ast}$ is an arbitrary pivot scale, $A_s$ is the scalar amplitude and $n_{\textrm{s}}$ is the tilted spectral index of scalar perturbations \cite{Adam:2014bub}. The parameters $A_s$ and $n_s$  which have been measured to high precision by Planck team are $A_{\textrm{s}}=2.2\times 10^{-9}$ \cite{Ade:2013zuv,Planck:2013jfk} and $n_{\textrm{s}}=0.968\pm 0.006$ \cite{Ade:2015lrj,Planck:2015xua}. An upper bound on the tensor-to-scalar ratio of $r_{0.05}< 0.12$ (95\% CL) \cite{Ade:2015lrj} has also been obtained from Planck in combination with BICEP2/Keck Array and BAO (Baryon acoustic oscillations) data. Recently, this upper bound on $r$ has been tightened to $r_{0.05}< 0.07$ (95\% CL) by BICEP2/Keck plus Planck data \cite{Array:2015xqh}. Based on these constraints, the simple models of chaotic potential are already ruled out and the Starobinsky model predicting a low value of $r$ is one of the most favored model.\\
\indent
After Planck 2013, a broad class of the inflationary models have been proposed based on the conformal symmetry in Jordan frame, all predicting a universal attractor behavior in the Einstein frame \cite{Kallosh:2013lkr,Kallosh:2013pby,Kallosh:2013hoa,
Ferrara:2013rsa,Kallosh:2013tua,Kallosh:2013yoa,Kallosh:2014rga,Kallosh:2014laa,Galante:2014ifa,Kallosh:2015lwa,Linde:2015uga,Carrasco:2015rva,Carrasco:2015pla}. The most interesting classes of such supergravity inflationary models are ``$\alpha$-attractors'' in which potentials involve a free parameter $\alpha$ \cite{Kallosh:2013yoa,Kallosh:2014rga,Kallosh:2014laa,Galante:2014ifa,Kallosh:2015lwa,Linde:2015uga,Carrasco:2015rva,Carrasco:2015pla}. For one of these classes known as T-models, the Einstein frame potential takes the form $f^2(\tanh(\varphi/(\sqrt{6\alpha}M_{\textrm{Pl}})))$ while for another interesting class called E-models, the potential is given by $f^2(1-\exp(-\sqrt{2/3\alpha}\varphi/M_{\textrm{Pl}}))$. Here $M_{\textrm{Pl}}$ is the reduced Planck mass. For large e-folding number $N_k$ and small $\alpha$, these models have the same predictions \cite{Kallosh:2013yoa,Kallosh:2014rga,Kallosh:2014laa,Galante:2014ifa,Kallosh:2015lwa,Linde:2015uga,Carrasco:2015rva,Carrasco:2015pla} corresponding to the central area of the $n_{\textrm{s}}-r$ plane favored by Planck 2015 \cite{Planck:2015xua}.\\
\indent
Although it is difficult to constrain reheating era observationally, but the existence of upper and lower bounds on the reheating temperature helps us study model dependent reheating parameters like temperature and number of e-folds as functions of the scalar spectral index. Consequently, using the precision measurement of the spectral index one can put additional constraints on the inflationary models to break the degeneracy of wide variety of the inflationary models with similar predictions for inflationary observables.\\
\indent
In this paper, we will study the reheating constraints on the physical predictions of the $\alpha$-attractor models using the methods developed in \cite{Liddle:2003as,Martin:2010,Creminelli:2014oaa,Dai:2014jja,Munoz:2014eqa,Creminelli:2014fca,Mielczarek,Martin:2014nya,Cook:2015vqa} to calculate the reheating e-folding number $N_{\textrm{re}}$ and reheating temperature $T_{\textrm{re}}$ as functions of $\alpha$ and $n_s$. Using the new CMB constraint on $n_{\textrm{s}}$ and $r$ and considering the lower bound on the $T_{\textrm{re}}$ given by primordial nucleosynthesis (BBN) and the upper bound on $T_{\textrm{re}}$ from the bound on $r$, we investigate the allowable range of the parameter $\alpha$. To do this precisely, we first compare the analytical slow-roll approximation to the exact numerical solution for the background equations and consequently take the unignorable correction terms into account during next reheating analysis. \\
\indent
The paper is organized as follows: In section \rom{2}, we review the main structure and predictions of the $\alpha$-attractor models, specially T- and E-models. Then, we check the precision of the cosmological predictions of these models via comparing of the slow-roll approach and the full numerical results. In section \rom{3}, we discuss the constraints of reheating calculations on the T-models and E-models. The paper concludes with a summary in section \rom{4}.

\section{Inflationary $\alpha\,$-attractor models}

 In last two years, several classes of the large field models called cosmological $\alpha$-attractor have been found with similar observational predictions. The Lagrangian of the $\alpha$-attractor models in the Einstein frame involving a real scalar field $\phi$ minimally coupled to gravity is given by \cite{Kallosh:2013yoa,Kallosh:2014rga,Kallosh:2014laa,Galante:2014ifa,Kallosh:2015lwa,Linde:2015uga,Carrasco:2015rva,Carrasco:2015pla,Roest:2015qya,Scalisi:2015qga}
\beq
\mathcal{L}=\sqrt{-g}\left[\frac{M_{\textrm{Pl}}^2}{2}R-\frac{\alpha}{(1-\phi^{2}/(3M_{\textrm{Pl}}^2))^{2}}(\partial\phi)^{2}-f^{2}(\dfrac{\phi}{\sqrt{3}M_{\textrm{Pl}}})\right]~,
\eeq
where $\alpha$ is a constant describing the inverse curvature of K{\"a}hler manifold and $f^2$ is an arbitrary function which plays the role of potential term. One can find inflationary solutions for all values of $\alpha$. Employing the canonical normalization of the kinetic term by the redefinition $\phi/\sqrt{3}=\tanh(\varphi/\sqrt{6\alpha})$, we arrive at the following class of $\alpha$-attractors known as T-models \cite{Kallosh:2013yoa,Kallosh:2014rga,Galante:2014ifa,Kallosh:2015lwa,Linde:2015uga,Carrasco:2015rva,Carrasco:2015pla}
\beq
\mathcal{L}=\sqrt{-g}\left[\frac{M_{\textrm{Pl}}^2}{2}R-\frac{1}{2}(\partial\varphi)^{2}-f^{2}\left(\tanh\frac{\varphi}{\sqrt{6\alpha}M_{\textrm{Pl}}}\right)\right]~.
\eeq
For this class with $f(x) \sim x^n$, the potential is given by \cite{Kallosh:2015lwa}
 \bea
V(\varphi)=\lambda_{n}\tanh ^{2n}\left(\dfrac{\varphi}{\sqrt{6\alpha}M_{\textrm{Pl}}}\right)~,\label{Tmodelpotential}
\eea
where $\lambda_{n}$ is a constant coefficient. This potential is symmetric with respect to $\varphi\rightarrow -\varphi$. \\
\indent

\begin{figure}[h]
\epsfig{figure=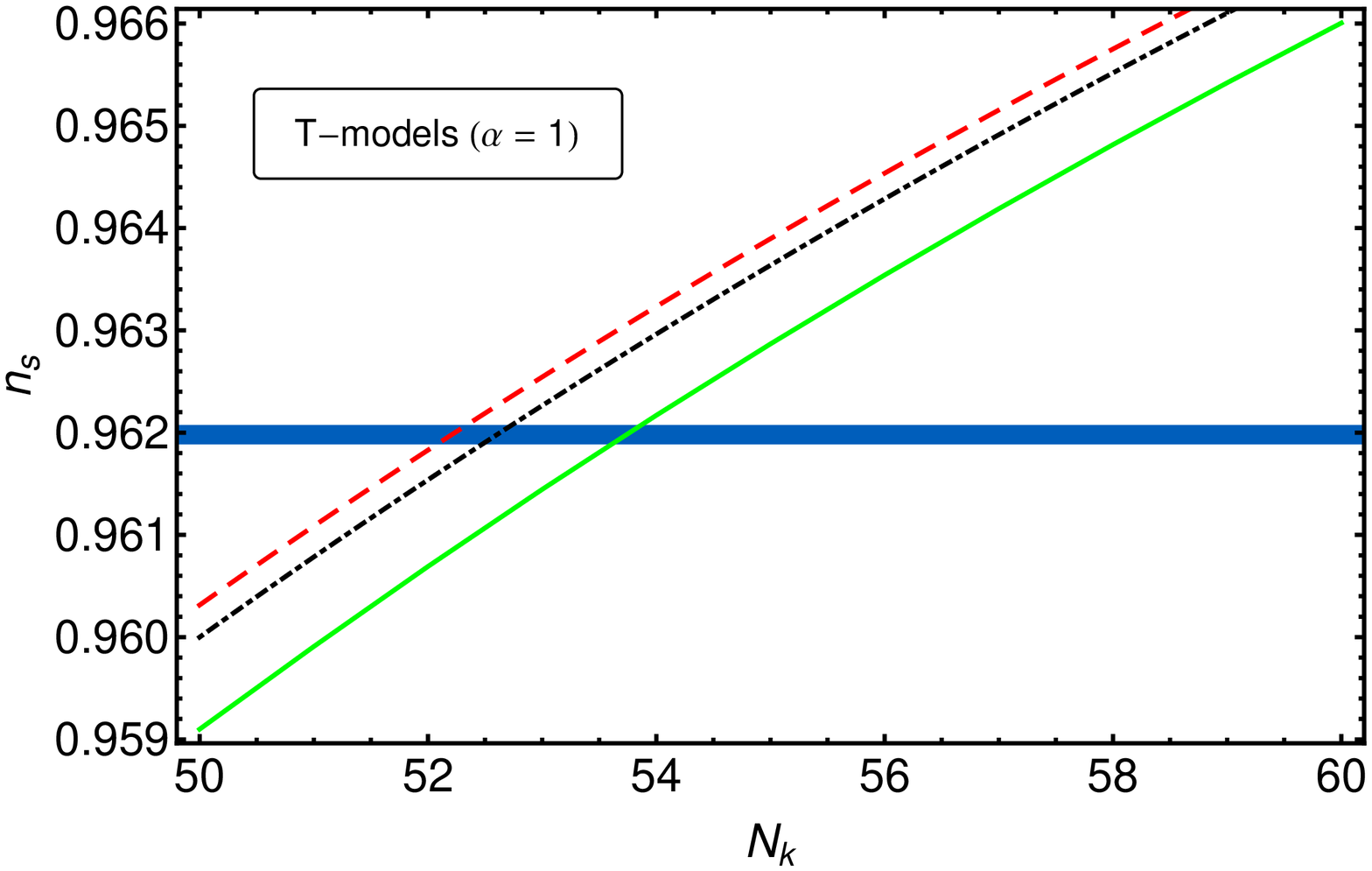,width=8.75cm}\hspace{2mm}
\epsfig{figure=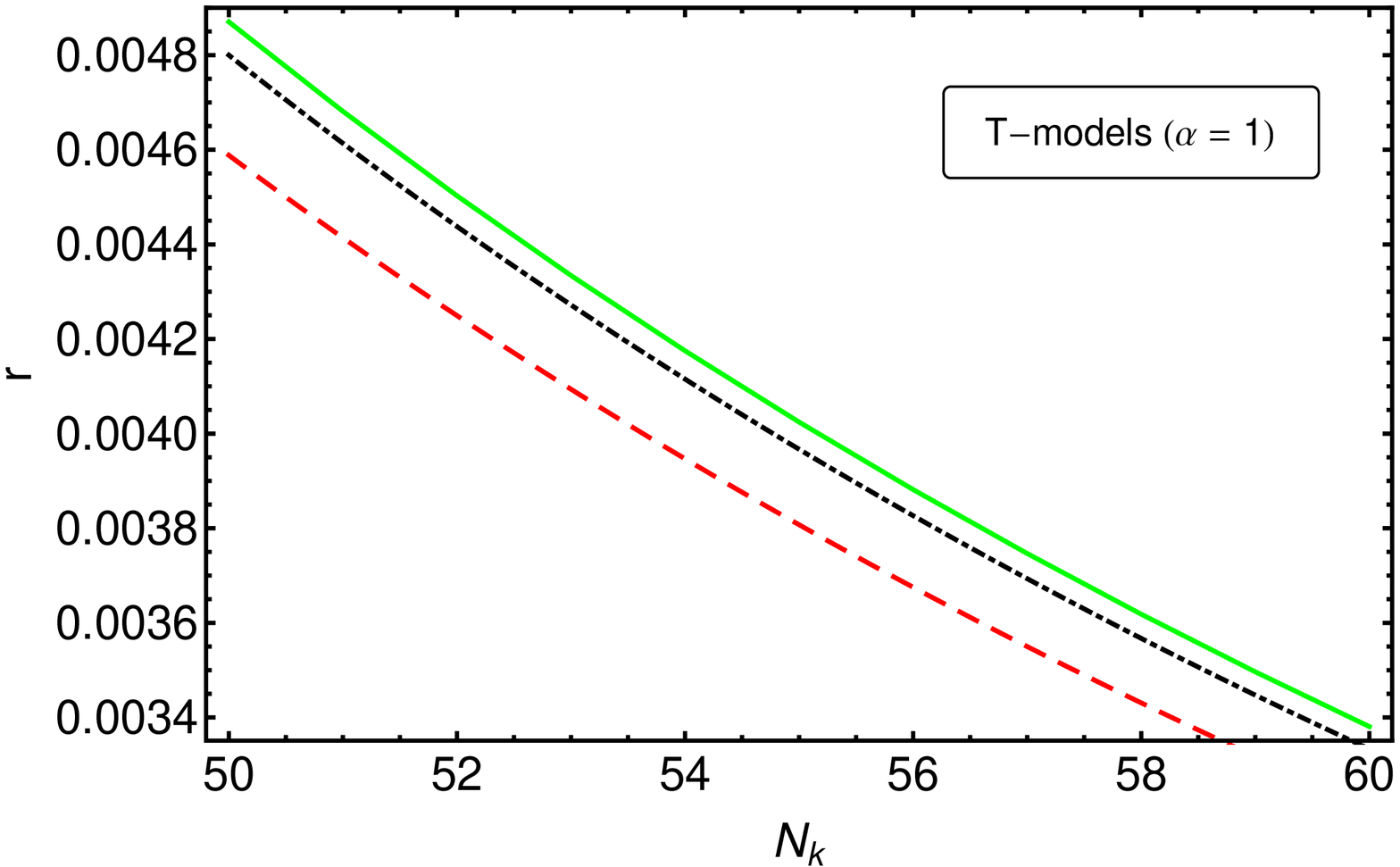,width=8.75cm}\hspace{2mm}
\epsfig{figure=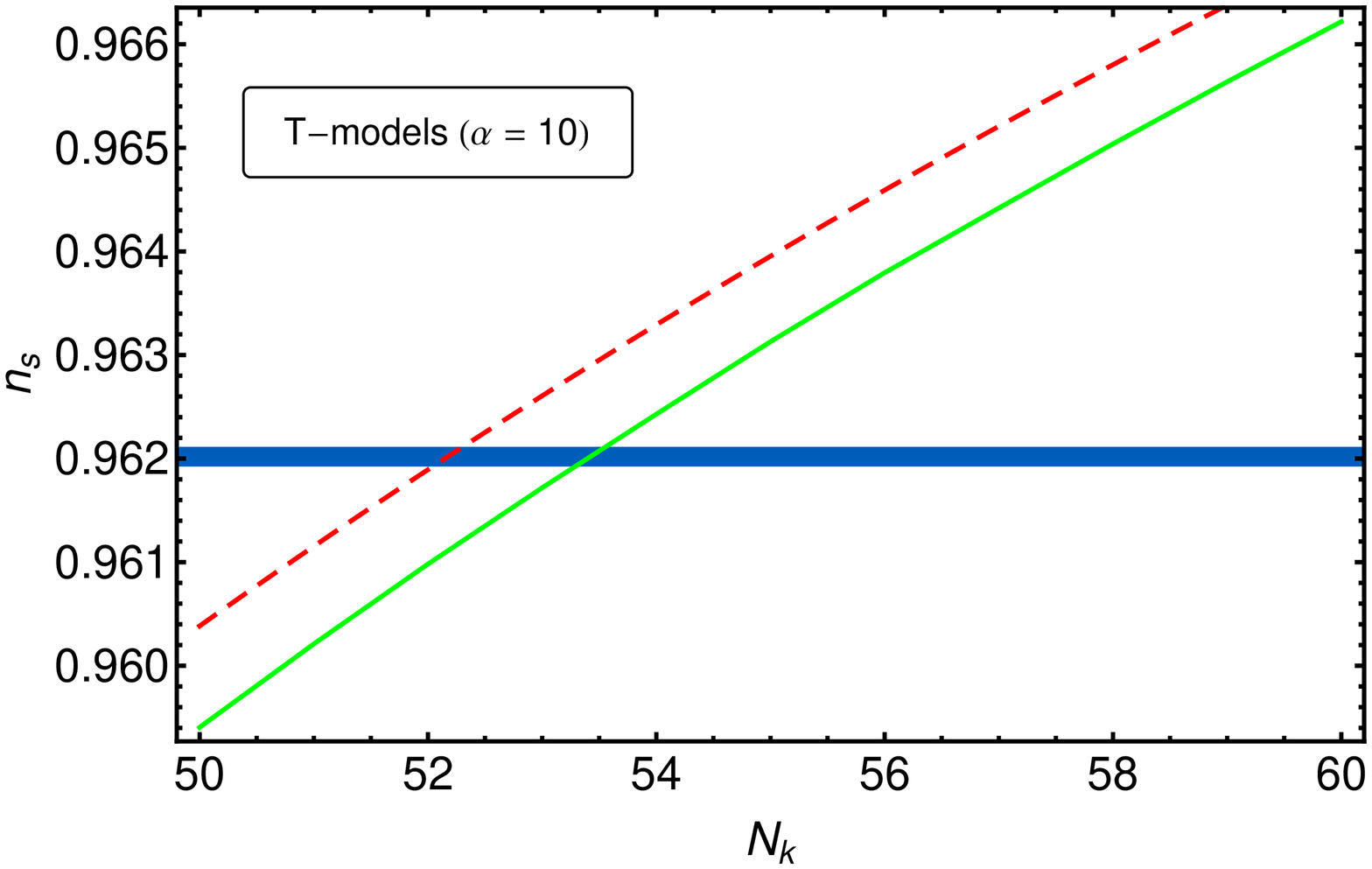,width=8.75cm}\hspace{2mm}
\epsfig{figure=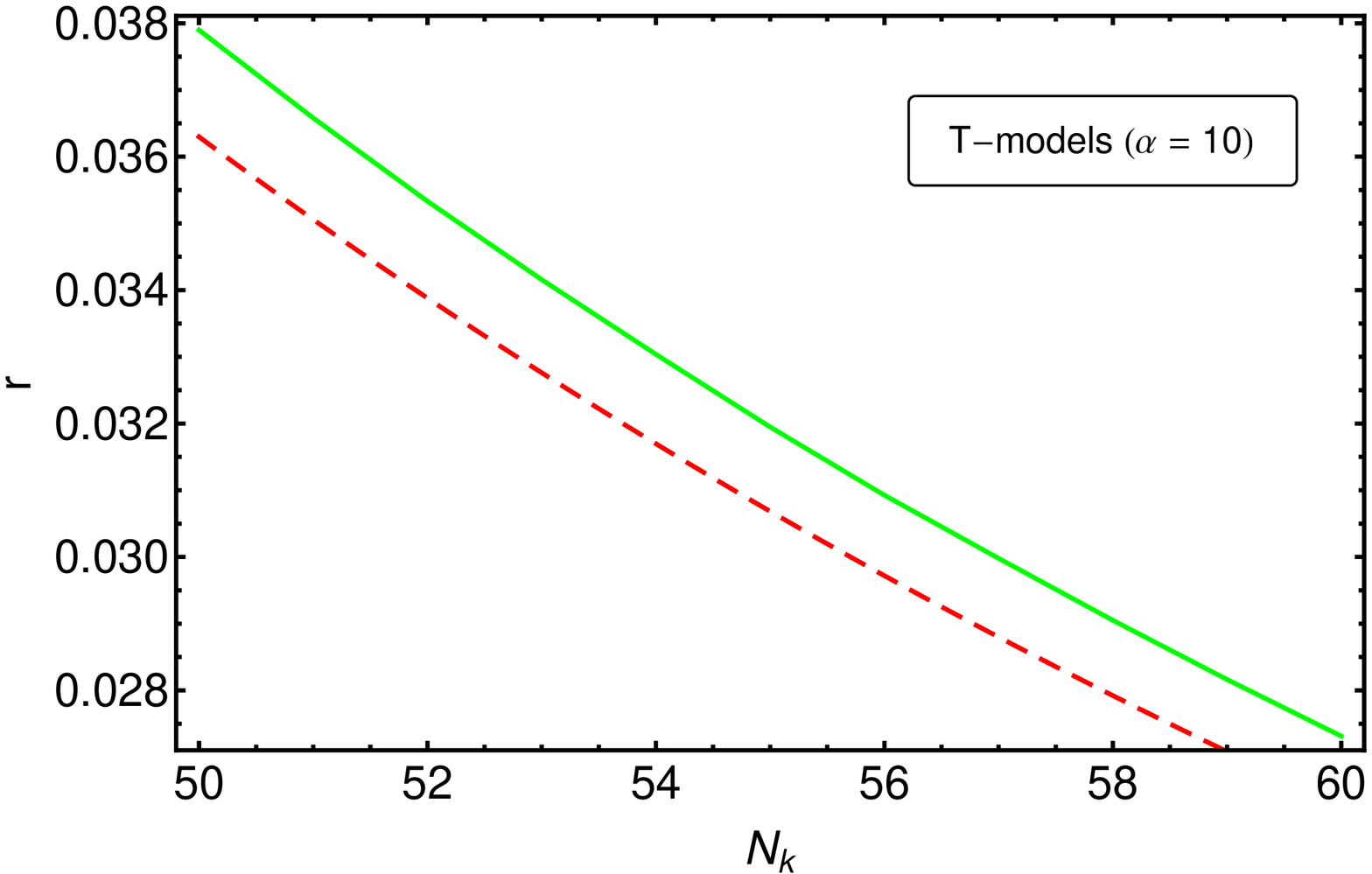,width=8.75cm}\hspace{2mm}
\epsfig{figure=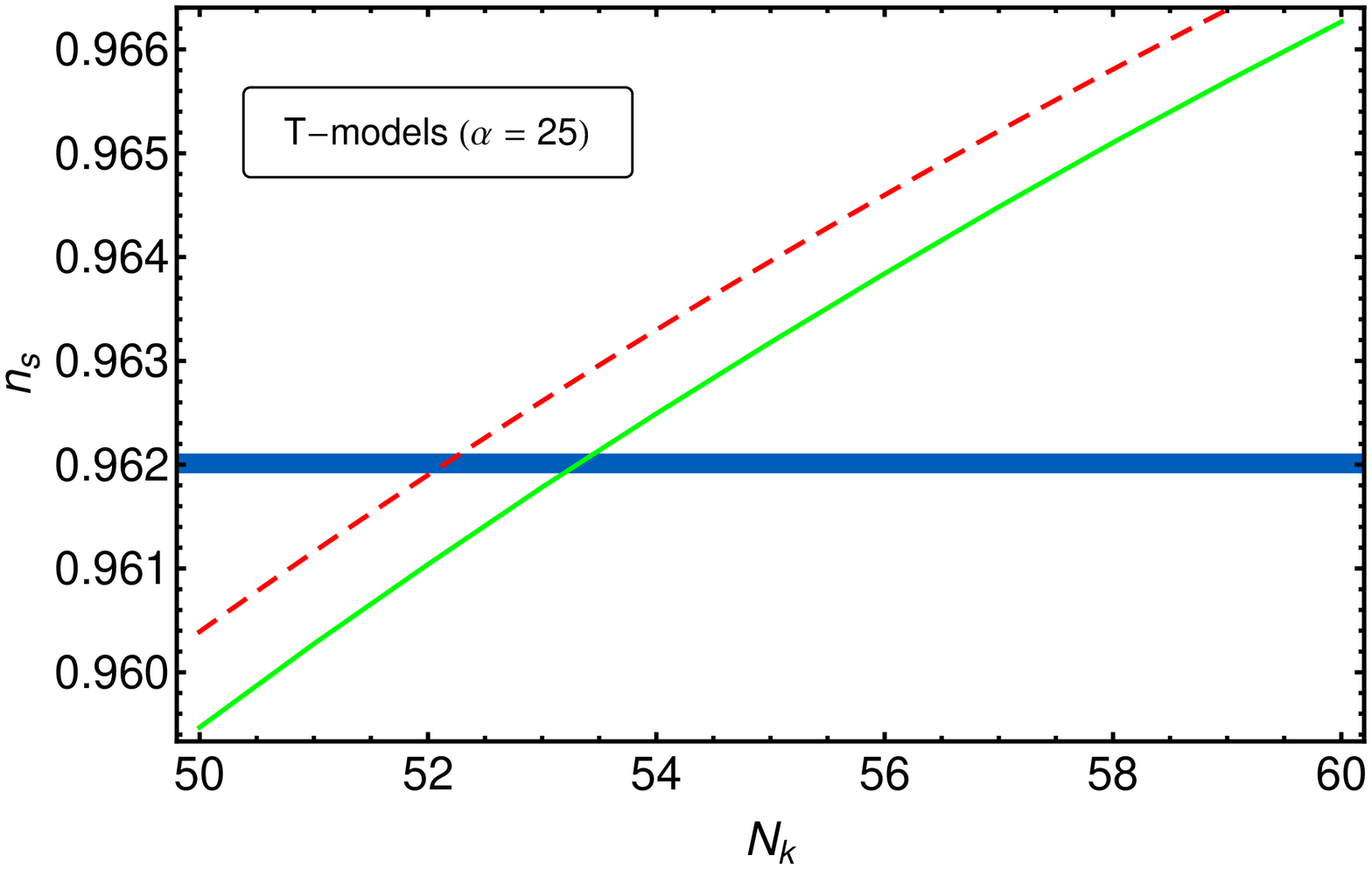,width=8.75cm}\hspace{2mm}
\epsfig{figure=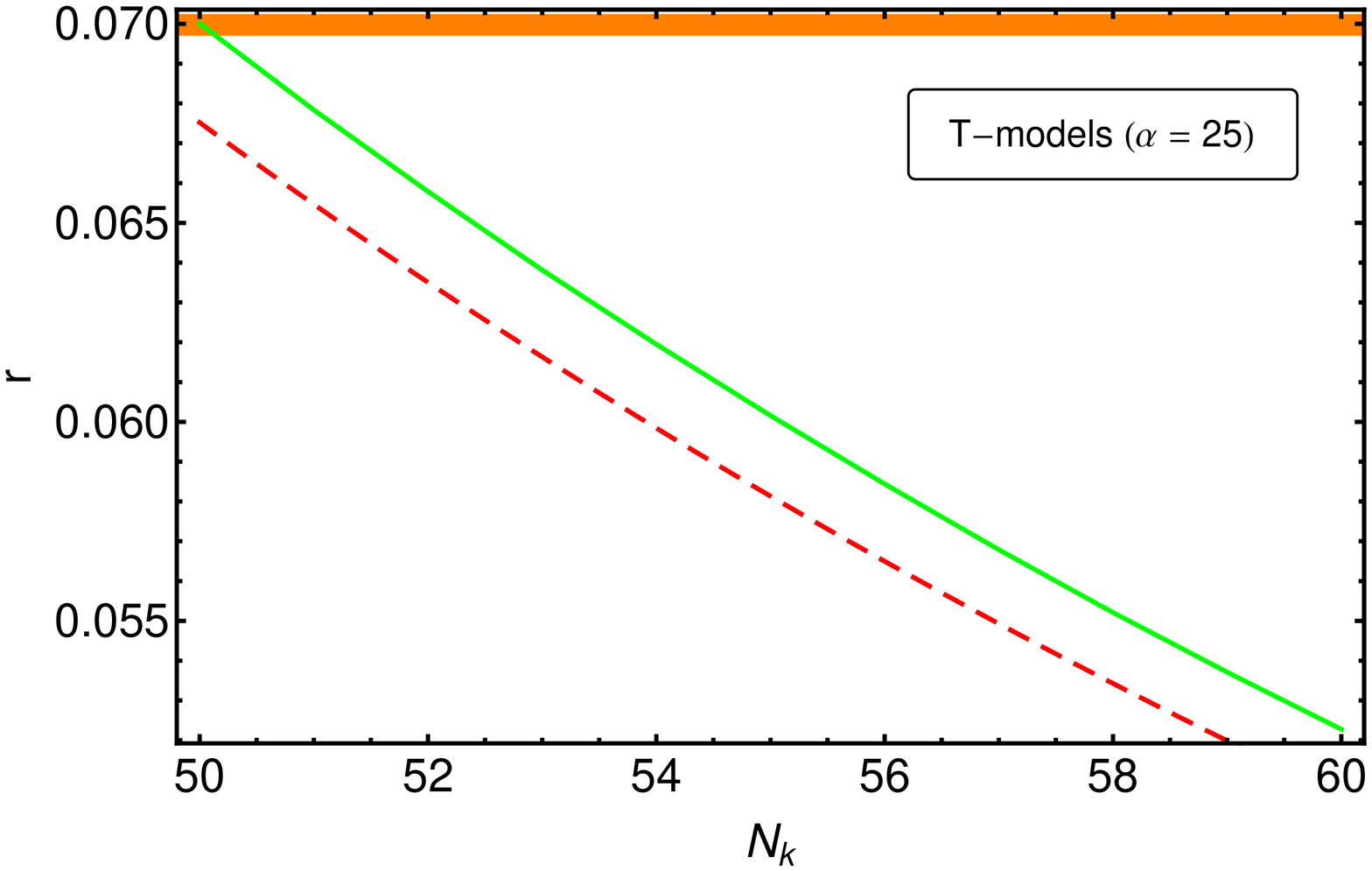,width=8.75cm}\hspace{2mm}
\caption{The spectral index, $n_s$  (left panels) and the tensor-to-scalar ratio, $r$ (right panels) as functions of the number of e-folds, $N_k$ for T-models with $n=1$. The exact numerical data are shown as green solid lines, while the universal prediction \eqref{123} and the slow-roll cosmological parameters \eqref{11} are displayed as the black dot-dashed and the red dashed lines, respectively. The horizontal dark blue lines show the 68\% CL lower limits from the Planck 2015 data and the orange one is due to the upper bound on $r< 0.07$ (95\% CL) by BICEP2/Keck plus Planck data.}
 \label{fig1}
\end{figure}

\begin{figure}[h]
\epsfig{figure=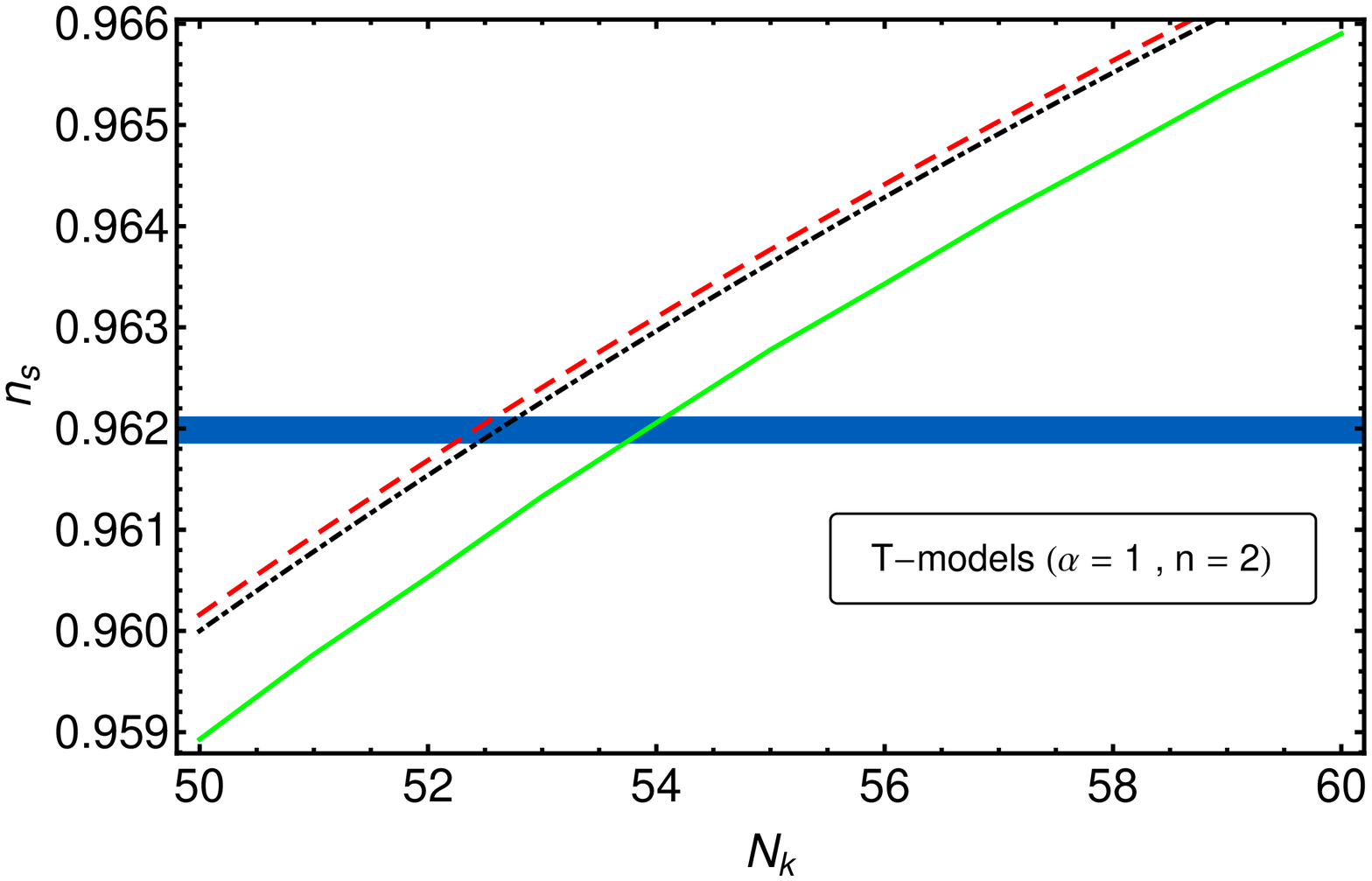,width=8.75cm}\hspace{2mm}
\epsfig{figure=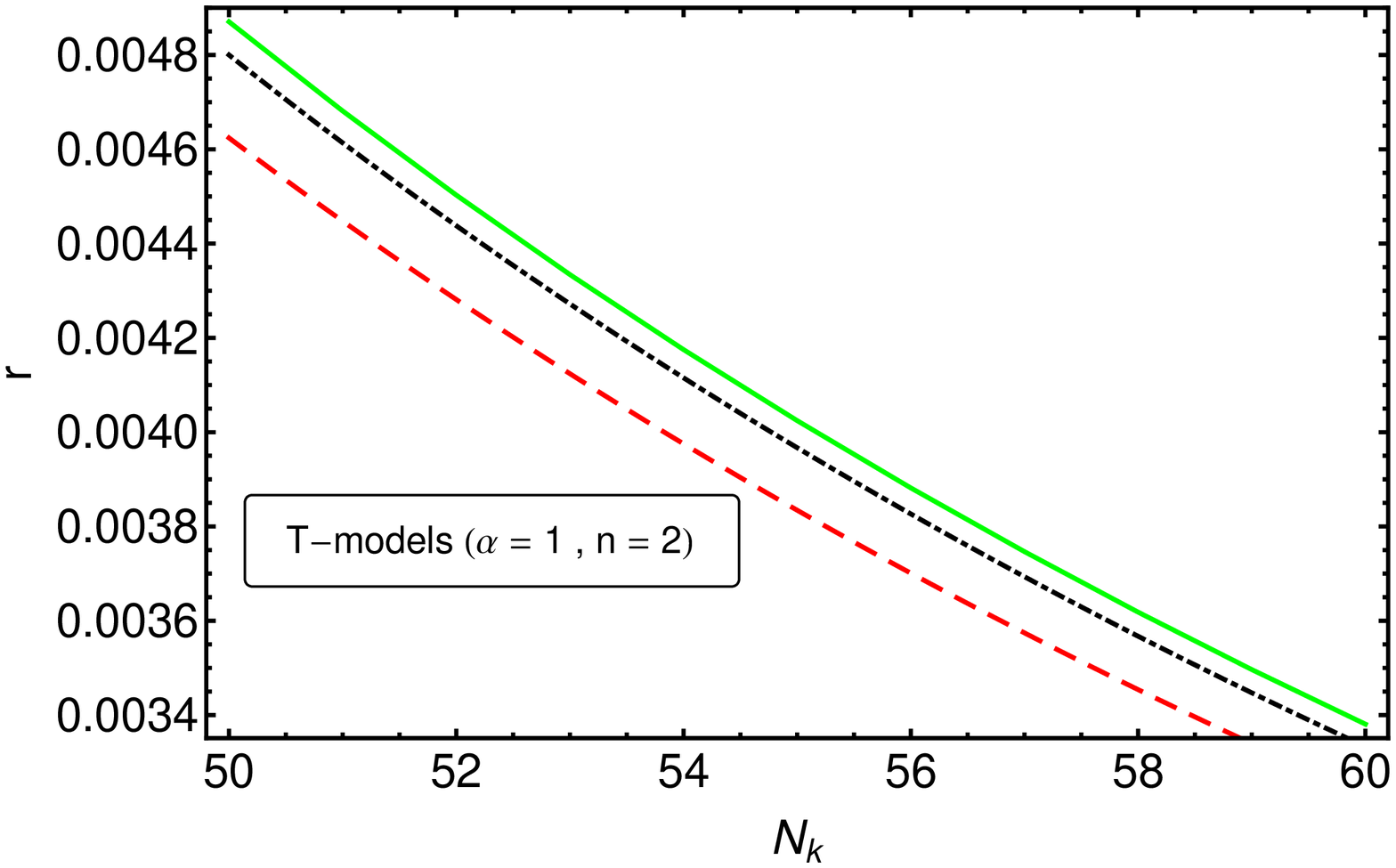,width=8.75cm}\hspace{2mm}
\epsfig{figure=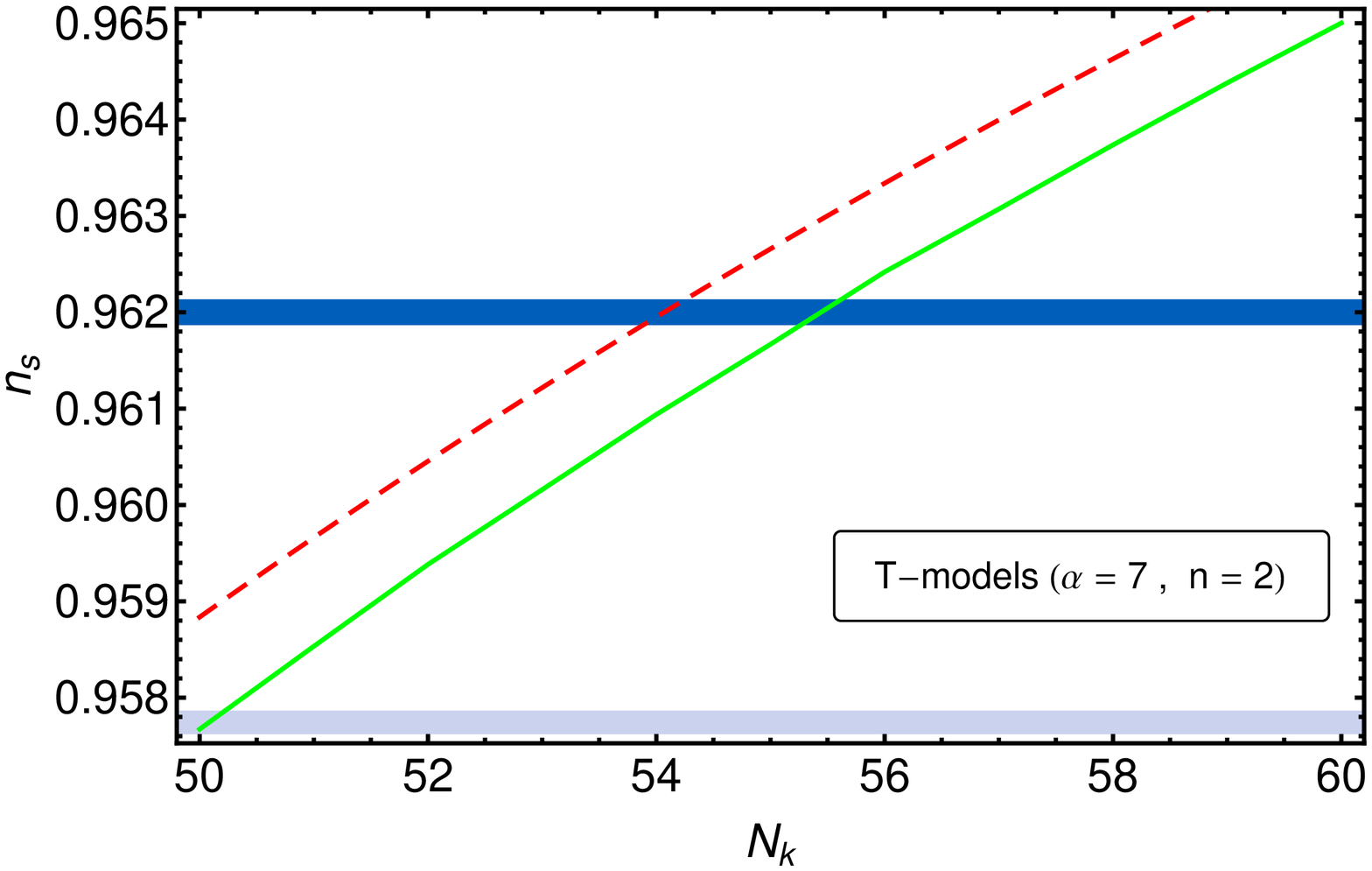,width=8.75cm}\hspace{2mm}
\epsfig{figure=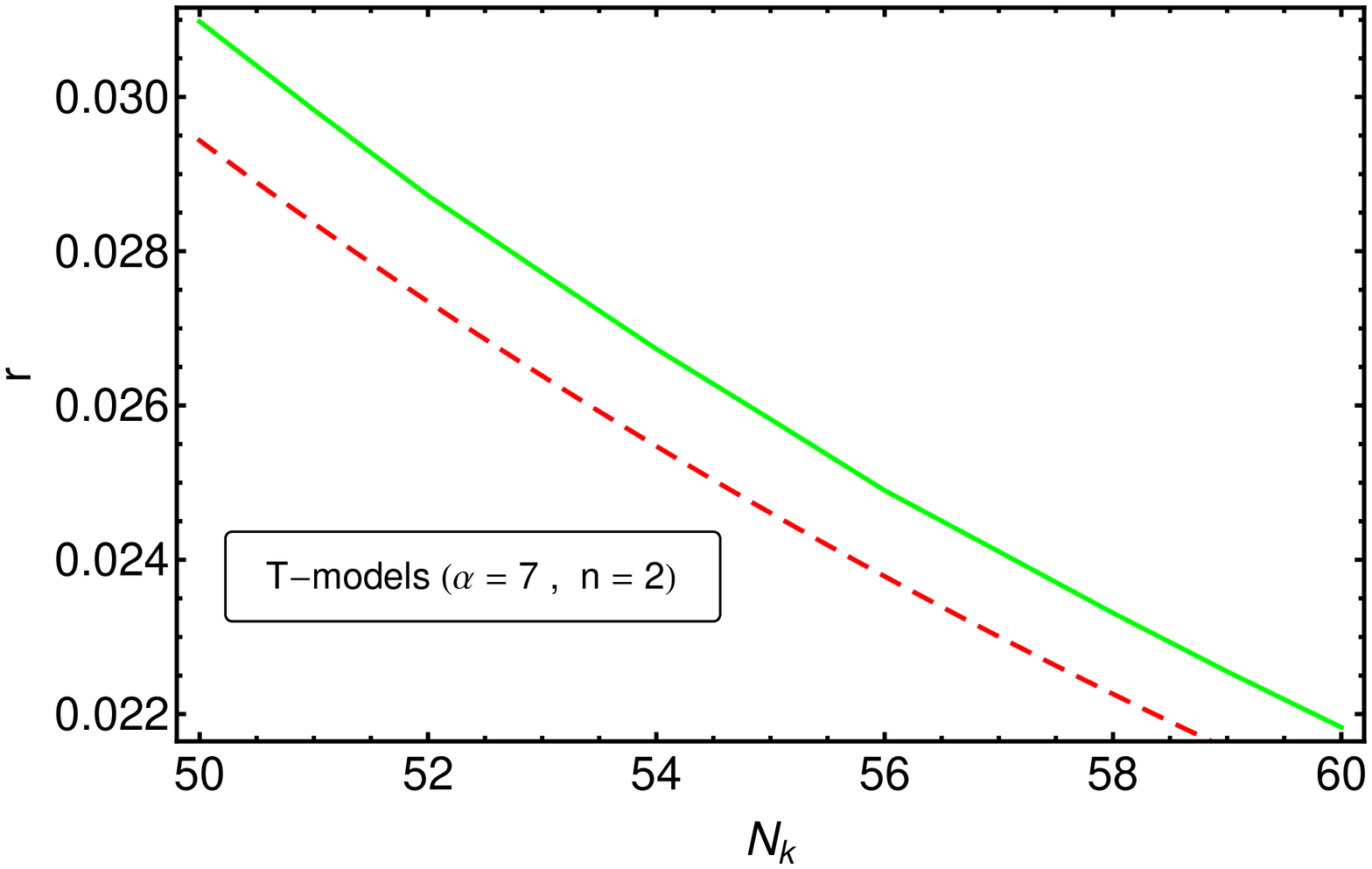,width=8.75cm}\hspace{2mm}
\caption{The spectral index, $n_s$  (left panels) and the tensor-to-scalar ratio, $r$ (right panels) as functions of the number of e-folds, $N_k$ for T-models with $n=2$. The exact numerical data are shown as green solid lines, while the universal prediction \eqref{123} and the slow-roll cosmological parameters \eqref{11} are displayed as the black dot-dashed and the red dashed lines, respectively. The horizontal dark blue and light blue lines show the 68 and 95\% CL lower limits from the Planck 2015 data.}
 \label{fig2}
\end{figure}

\begin{figure}[h]
\epsfig{figure=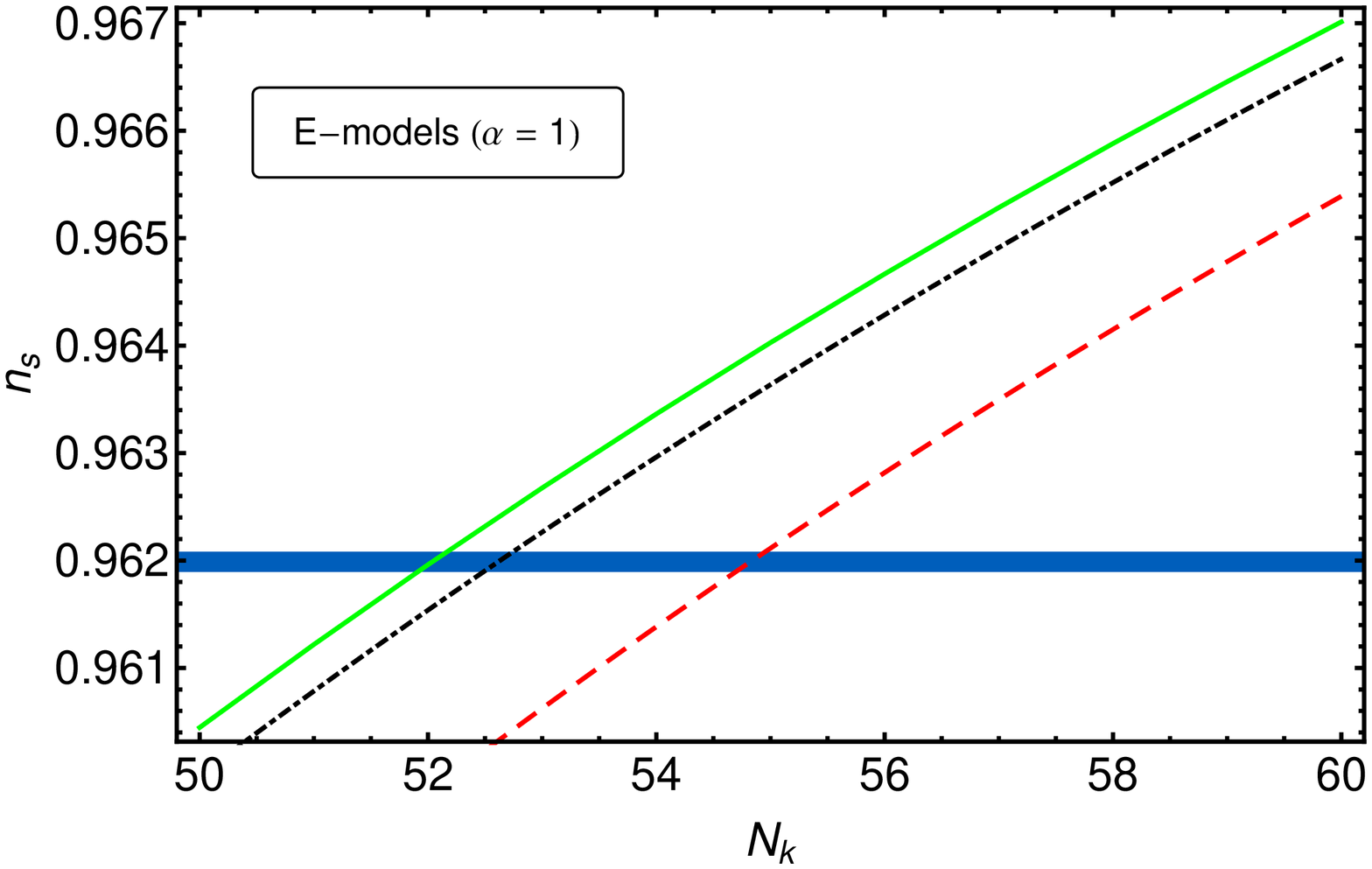,width=8.75cm}\hspace{2mm}
\epsfig{figure=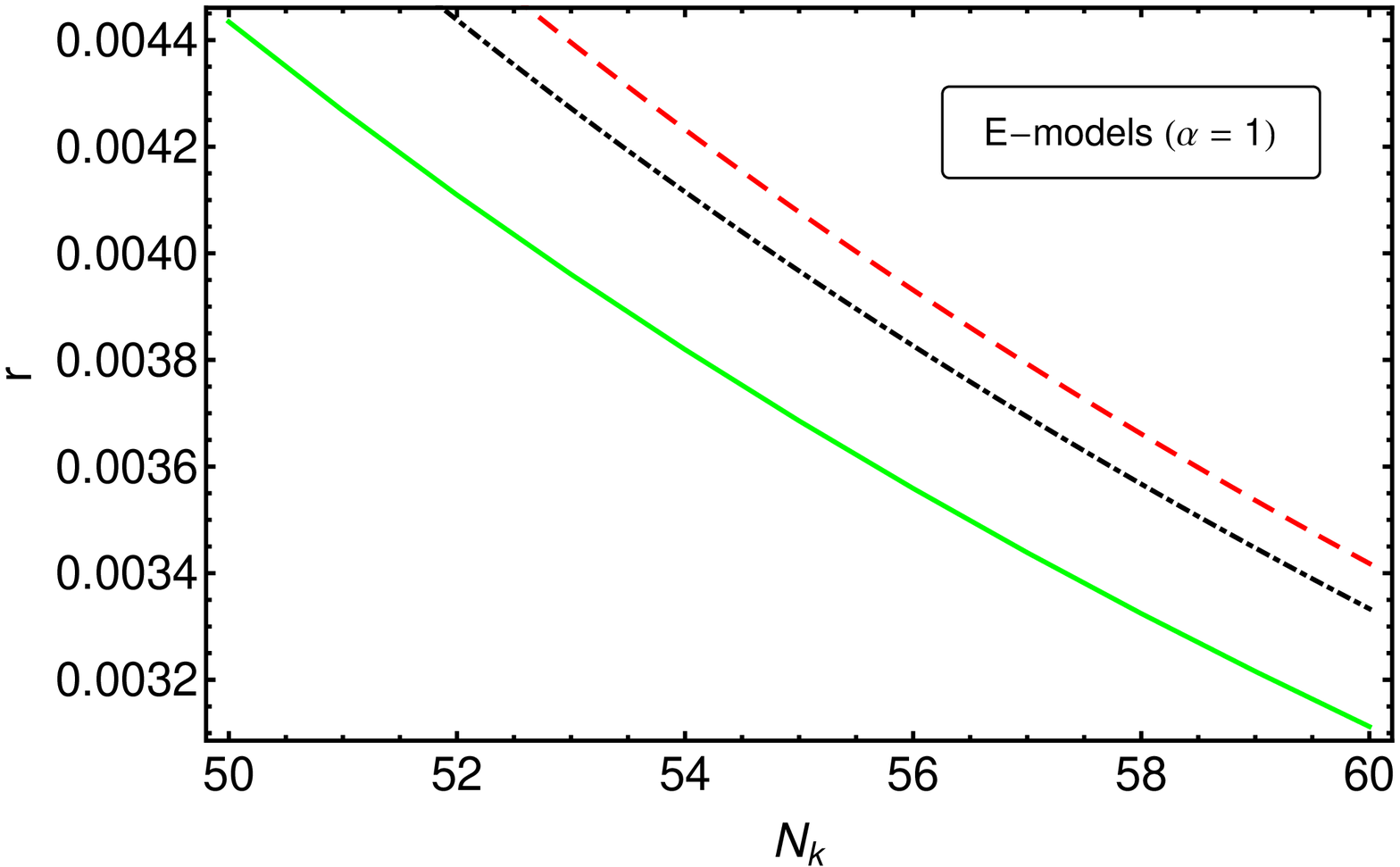,width=8.75cm}\hspace{2mm}
\epsfig{figure=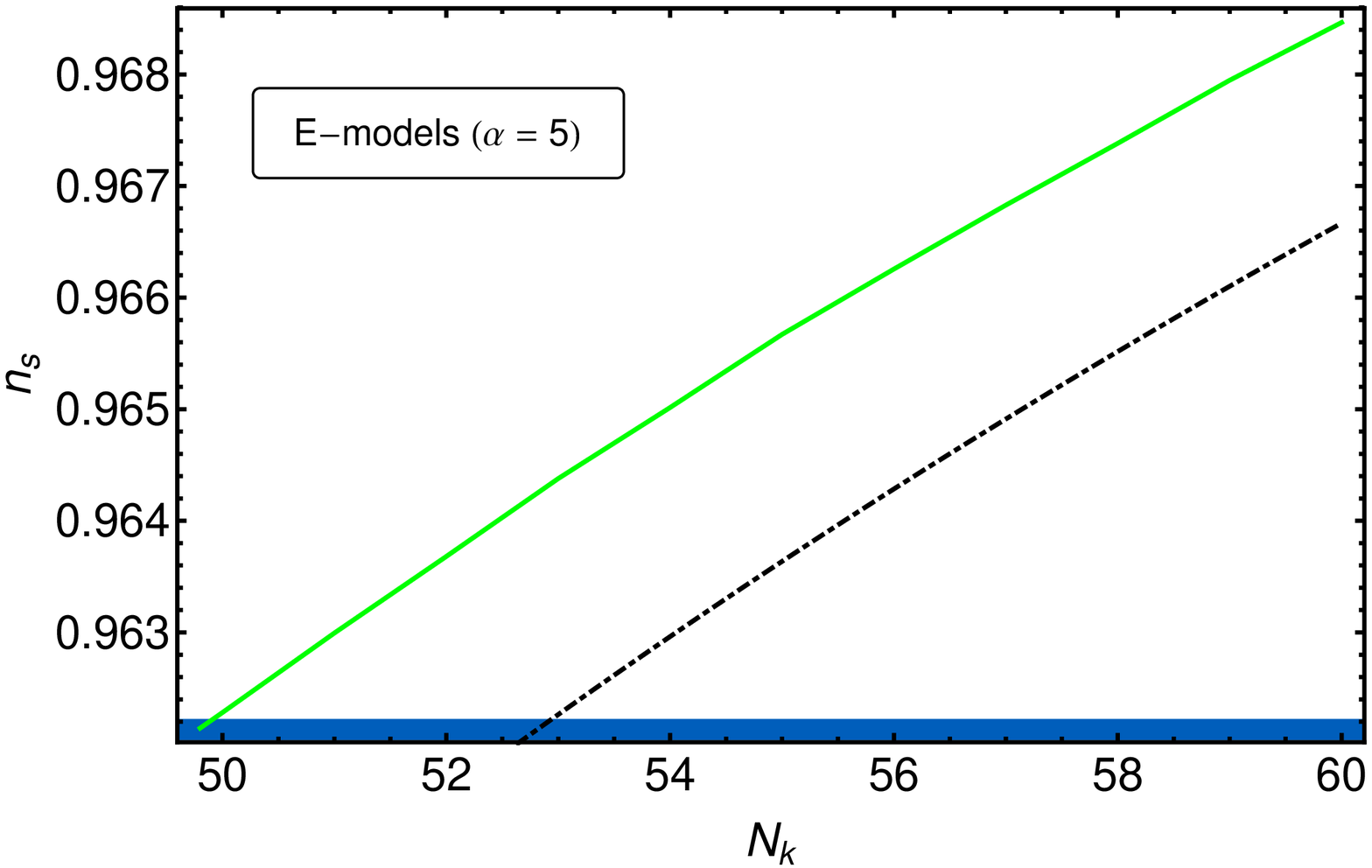,width=8.75cm}\hspace{2mm}
\epsfig{figure=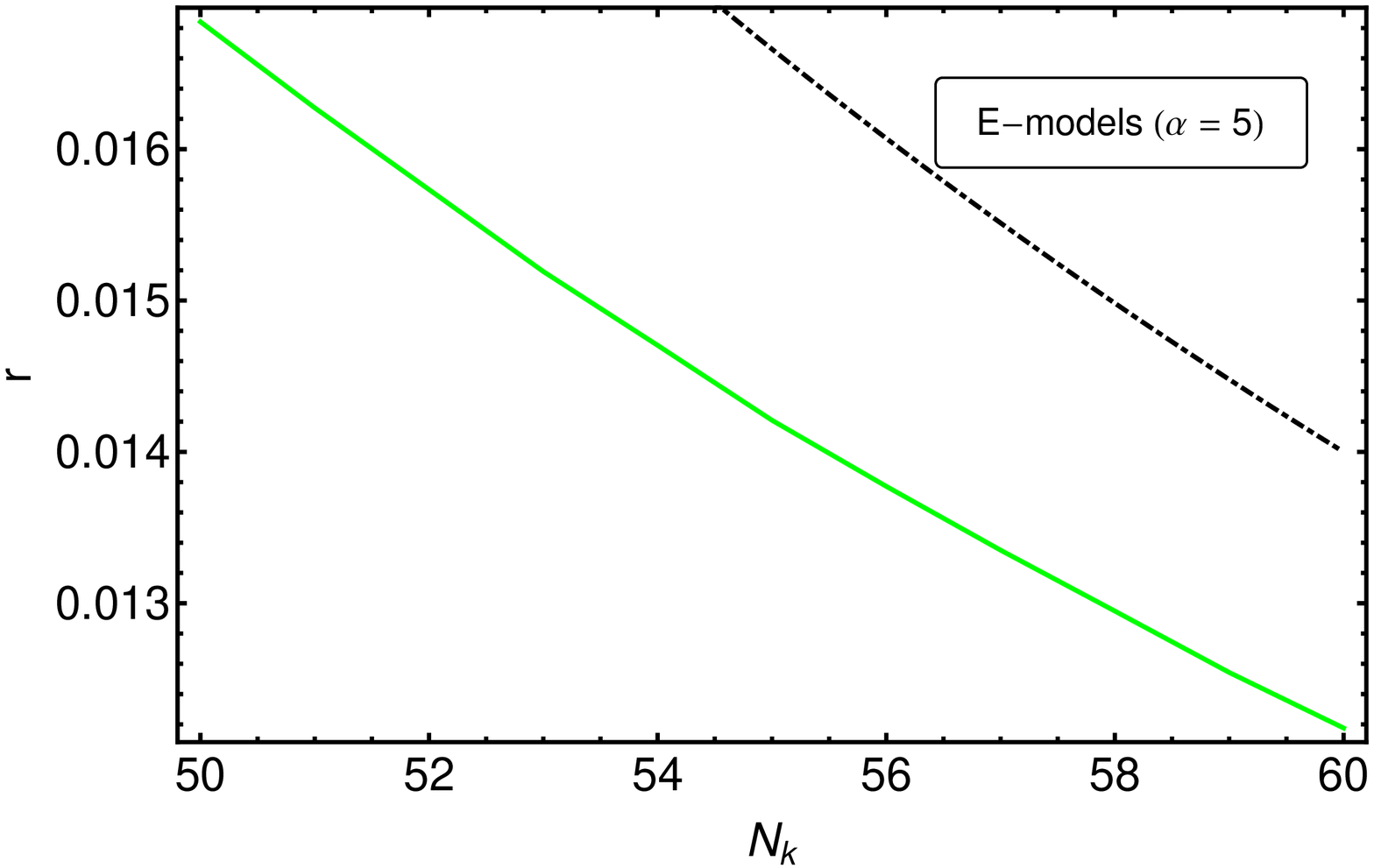,width=8.75cm}\hspace{2mm}
\epsfig{figure=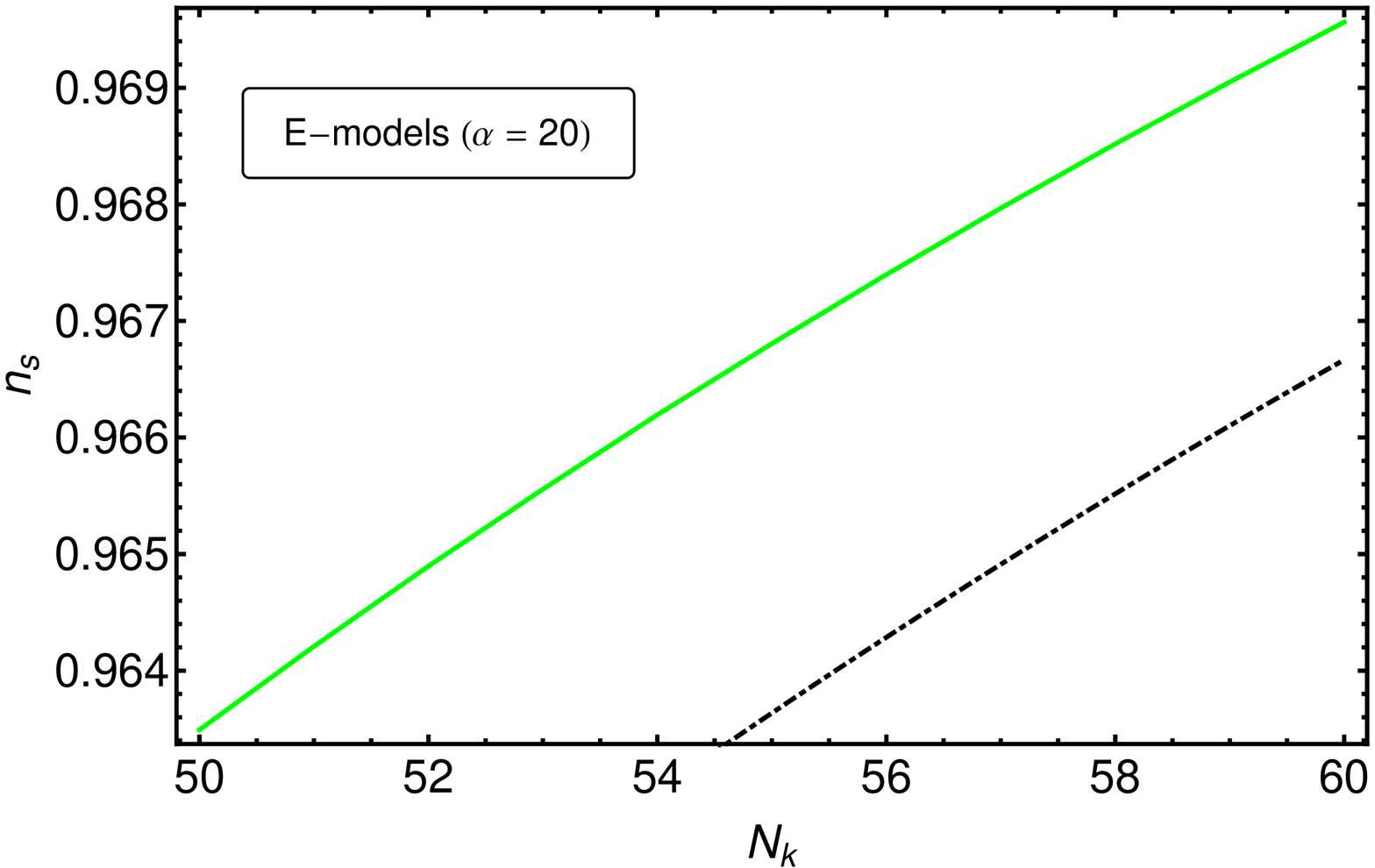,width=8.75cm}\hspace{2mm}
\epsfig{figure=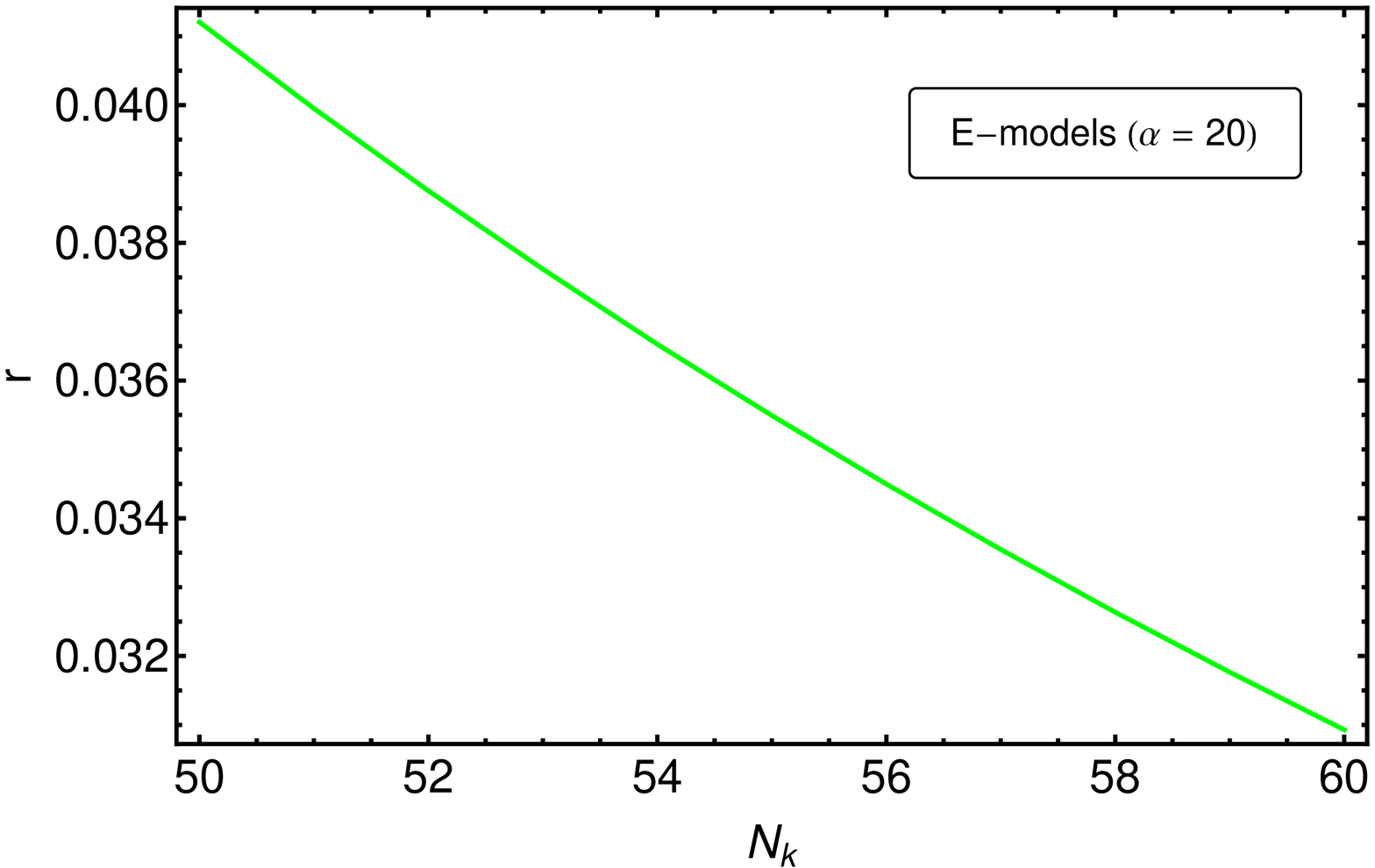,width=8.75cm}\hspace{2mm}
\caption{The spectral index, $n_s$  (left panels) and the tensor-to-scalar ratio, $r$ (right panels) as functions of the number of e-folds, $N_k$ for E-models. The exact numerical data are shown as green solid lines, while the universal prediction \eqref{123} and the slow-roll cosmological parameters \eqref{111} are displayed as the black dot-dashed and the red dashed lines, respectively. The horizontal blue lines show the 68\% CL lower limits from the Planck 2015 data, respectively.}
 \label{fig3}
\end{figure}

Assuming slow-roll condition in the inflation era, the number of e-folds $N_k$ can be calculated by
\beq
 N_k \simeq \frac{1}{M_{\textrm{Pl}}^{2}}\int^{\varphi_k}_{\varphi_{\textrm{end}}}\frac{V}{V'}d\varphi~, \label{18}
\eeq
where $\varphi_{k}$ and $\varphi_{\textrm{end}}$ are the values of inflaton field when the pivot scale, k, exits the horizon
and when the inflation ends, respectively. The prime denotes differentiation with respect to $\varphi$. Once the form of inflationary potential is specified, one can invert \eqref{18} to find $\varphi_{k}$ as a function of $N_k$ and $\varphi_{\textrm{end}}$.
At the first order in slow-roll approximation, the scalar spectral index $n_{\textrm{s}}$ and the tensor to scalar ratio $r$ are defined as
\bea
n_{\textrm{s}}=1-6\epsilon_V(\varphi_{k})+2\eta_V(\varphi_{k})~,\:\:\:\:\:\:\textrm{and}\:\:\:\:\:\:\:\: r=16\epsilon_V(\varphi_{k})~,
\label{181}
\eea
where $\epsilon_V(\varphi)$ and $\eta_V(\varphi)$ are potential slow-roll parameters
\bea
  \epsilon_V(\varphi) = \frac{M_{\textrm{Pl}}^{2}}{2}\left(\frac{V'(\varphi)}{V(\varphi)}\right)^{2}~,\:\:\:\:\:\:\:\:\:\:\:\:\:\: \eta_V(\varphi) =M_{\textrm{Pl}}^{2}\frac{V''(\varphi)}{V(\varphi)}~.\label{182}
\eea
Therefore, in the regime $\alpha >1/3$ (the stability condition of the model) and $n \geq 1/2$ by using \eqref{18} and \eqref{181}, the cosmological predictions of potential \eqref{Tmodelpotential} are given by \cite{Kallosh:2013yoa}
\bea
 n_{\textrm{s}}(\alpha, n ,N_k) = \frac{1-\frac{2}{N_k}-\frac{3\alpha}{4N_k^2}+\frac{1}{2nN_k}(1-\frac{1}{N_k})g(\alpha, n)}{1+\frac{1}{2nN_k}g(\alpha, n)+\frac{3\alpha}{4N_k^2}}\: ,\:\:\:\:\:\:\:\:\:\:\:\:\:\:r(\alpha, n ,N_k) = \frac{12\alpha}{N_k^2+\frac{N_k}{2n}g(\alpha, n)+\frac{3}{4}\alpha}~, \label{11}
\eea
where $g(\alpha, n)=\sqrt{3\alpha(4n^2+3\alpha)}$. For $\alpha\lesssim O(1)$, the predictions of this class of models is unique for a broad set of choices of $f$.
For $\alpha=1$, $n_{\textrm{s}}$ and $r$ parameters predicted by this model coincide with the corresponding expressions of Starobinsky potential to the leading order of $1/N_k$ while in the limit $\alpha\rightarrow \infty$, this model predicts the same $n_{\textrm{s}}$ and $r$ parameters as $V \sim \varphi^{2n}$. One can also show an interpolation between the chaotic inflation potentials $V \sim \varphi^{2n}$ at large $\alpha$ and the universal attractor result at small $\alpha$. The comparison of this interpolation with the region in the $n_{\textrm{s}}-r$ plane predicted by recent data shows a good agreement between CMB data and T-models \cite{Kallosh:2013yoa,Kallosh:2015lwa}. \\
\indent
Another class of the $\alpha$-attractors called E-models is given by the following effective Lagrangian in Einstein frame \cite{Kallosh:2013yoa,Kallosh:2014rga,Galante:2014ifa,Kallosh:2015lwa,Linde:2015uga,Carrasco:2015rva,Carrasco:2015pla}
\bea
\mathcal{L}=\sqrt{-g}\left[\frac{M_{\textrm{Pl}}^2}{2}R-\dfrac{1}{2}(\partial\varphi)^2-f^2\left(1- e^{{-\sqrt{\frac{2}{3\alpha}}}\varphi/M_{\textrm{Pl}}}\right)\right]~, \label{4}
\eea
where in the case $f(x) \sim x^n$ and one can find the potential term of this model as \cite{Carrasco:2015pla}
\bea
V(\varphi)= \mu_{n} \left(1- e^{-\sqrt{\frac{2}{3\alpha}}\varphi/M_{\textrm{Pl}}}\right)^{2n}~,\label{5}
\eea
 in which $\mu_{n} $ is a constant corresponding to the energy scale of inflation.  For $\alpha=1$ and $n=1$, \eqref{5} gives the potential of the Starobinsky model \cite{Starobinsky:1980te}. Using \eqref{18} and \eqref{181}, the cosmological predictions of \eqref{5} are
\bea
 n_s(\alpha, n ,N_k) = 1-\dfrac{8n^2(4N_k+3\alpha)}{(4n N_k-3\alpha)^2} \: ,\:\:\:\:\:\:\:\:\:\:\:\:\:\:r(\alpha, n ,N_k) =\dfrac{192\alpha n^2}{(4n N_k-3\alpha)^2}~. \label{111}
\eea
Comparing with Planck 2013 and Planck 2015 data on $n_s-r$ plane, one concludes that the potential \eqref{5} with small $\alpha$  favors the data very well \cite{Kallosh:2013yoa,Kallosh:2015lwa}. As one can see for $N_k\gg 1$ both of T- and E-models have the same universal predictions
\cite{Ferrara:2013rsa,Kallosh:2013yoa}
\beq
n_s=1-\frac{2}{N_k}~,\:\:\:\:\:\:\:\:\:\:\:\:\:\:\:r=\frac{12\alpha}{N_k^2}~. \label{123}
\eeq
Before going through the reheating analysis of these models, in the rest of this section we test the precision of results \eqref{123} which is an important ingredient for our next calculations.\\
\indent
During inflation the dynamics of inflaton field is governed by the Klein-Gordon and Friedmann equations respectively
 \bea
 \ddot{\varphi}+3H\dot{\varphi}+V'=0~,\:\:\:\:\:\:\:\:\:\:\:\:\:\: \ H^{2} = \frac{1}{3M_{\textrm{Pl}}^{2}}\left(\frac{\dot{\varphi}^{2}}{2}+V\right)^{2} ,\label{183}
 \eea
where the dot denotes derivative with respect to time. The numerical integration of these equations between the horizon crossing and the end of inflation gives exact $n_s(N_k)$ for both classes of $\alpha$-attractor. Then, comparing this numerical answer with the results \eqref{123} given by slow roll approximation, the universal expression for $n_s(N_k)$ and $r(N_k)$ will give the corrections in the following forms
\beq
n_s=1-\frac{2}{N_k}+\delta~,\:\:\:\:\:\:\:\:\:\:\:\:\:\:\:r=\frac{12\alpha}{N_k^2}+\lambda~ \label{184},
\eeq
where $\delta$ and $\lambda$ have been adopted to make more accurate predictions for $n_s(N_k)$ and $r(N_k)$ with respect to the slow-roll approximation. For instance, for both T- and E-models we integrate the equations \eqref{183} numerically and compute the values of $n_{\textrm{s}}$ and $r$ some $50$-$ 60$ e-folds before the end of inflation given by zero acceleration. For $\alpha=1$ and $n=1$, comparison of the numerical values of $n_{\textrm{s}}(N_k)$ and $r(N_k)$ with \eqref{184} yields $\delta=-0.0009$ and $\lambda = 0.00005$ for T-models and $\delta=0.0004$ and $\lambda= -0.0003$ for E-models (i.e. the Starobinsky model).  Figs.  \ref{fig1}-\ref{fig3} show the results of $n_{\textrm{s}}(N_k)$ and $r(N_k)$  for T- and E-models.

Fig. \ref{fig1} shows that the simplest T-models with $n=1$ are vulnerable for $N_k\lesssim54$ since they leads to $n_s$ which is out of $1\sigma$ confidence region from the Planck 2015 data although it is still in $2\sigma$ confidence region. On the other hand, considering the upper bound on the tensor-to-scalar ratio $r< 0.07$ (95\% CL) by BICEP2/Keck plus Planck data \cite{Array:2015xqh}, this figure indicates that T-models with $n=1$ and $\alpha>25$ are unfavorable.\\
\indent
Fig. \ref{fig2} displays the behavior of $n_s$ and $r$ for T-models with $n=2$. In this case, $N_k\lesssim53.5$ leads to $n_s$ which is out of $1\sigma$ region but it is still in $2\sigma$ region. On the other hand, the maximum allowable value for $\alpha$ which keeps $n_s$ within 95\% CL is $\alpha=7$.\\
\indent
From Fig. \ref{fig3}, the starobinsky model ($\alpha=1$) is a bit unfavorable for $N_k<52$ as it is out of $1\sigma$ region, but for $\alpha\geq 5$, E-models are very safe as compared to Planck data (68\% CL). However, $r$ varies very slowly with $\alpha$ for E-models.

In terms of Hubble slow-roll parameters
\begin{equation}
\epsilon_H(\varphi) = 2M_{\textrm{Pl}}^{2}\left(\frac{H'(\varphi)}{H(\varphi)}\right)^{2}~,\:\:\:\:\:\:\:\:\:\:\:\:\:\:  \eta_H(\varphi) = 2M_{\textrm{Pl}}^{2}\frac{H''(\varphi)}{H(\varphi)},
\end{equation}
which are very small during inflation, inflation ends when $\epsilon_H(\varphi)=1$, i.e. zero acceleration, while if one considers potential slow-roll parameters \eqref{182}, then $\epsilon_V(\varphi)=1$ gives only a first order approximation for $\varphi_{\textrm{end}}$. Therefore, to find a more exact value of $\varphi_{\textrm{end}}$, one should work with a higher order approximation through \cite{Ellis:2015pla}
\beq
\epsilon_V(\varphi_{\textrm{end}})\simeq(1+\sqrt{1-\eta_V(\varphi_{\textrm{end}})/2})^2~. \label{185}
\eeq
It is also possible to obtain the most exact value for $\varphi_{\textrm{end}}$, and consequently for $V_{\textrm{end}}$, by using the numerical integration results of \eqref{183}.

\section{Reheating Analysis}

The main aim of this paper is to find the reheating constraints to T- and E-models which were considered in previous section. In order to recognize the relationship between inflation and reheating parameters, one may consider the connection between the time of horizon crossing of the cosmological observable scales and the time of their re-entering to the Hubble horizon \cite{Liddle:2003as}
\bea
\frac{k}{a_0H_0}=\frac{a_kH_k}{a_0H_0}=e^{-N_{k}}\frac{a_{end}}{a_{re}}\frac{a_{re}}{a_{eq}} \frac{H_{k}}{H_{\textrm{eq}}}\frac{a_{\textrm{eq}}H_{\textrm{eq}}}{a_0H_0}~ ,\label{13}
\eea
where the comoving wave number $k$ equals the Hubble scale $a_k H_k$  and the subscripts refer to different eras, including the horizon exit ($k$), reheating (re), radiation-matter equality (eq) and the present time (0). Accepting the assumption of entropy conservation between the end of reheating and today and using the slow-roll approximation in which $H^{2}_{k}\simeq V_k/3 M_{\textrm{Pl}}^2$ one can obtain \citep{Ade:2015lrj, Liddle:2003as,Martin:2010}
\bea
 N_k=66.9- \ln \left(\frac{k}{a_0 H_0}\right)+\frac{1}{4}\ln \left(\frac{V_k^2}{M_{\textrm{Pl}}^4}\rho_{\textrm{end}}\right)
 +\frac{1-3\omega_{\textrm{int}}}{12(1+\omega_{\textrm{int}})}\ln\left(\frac{\rho_{\textrm{re}}}{\rho_{\textrm{end}}}\right)-\frac{1}{12}\ln g_{\textrm{re}} ~,\label{16}
\eea
where $V_k$ is the potential energy when $k$ leaves the Hubble horizon during inflation, $\rho_{\textrm{end}}$ and $\rho_{\textrm{re}}$ are the energy densities at the end of inflation and reheating, respectively, $\omega_{\textrm{int}}$ is the e-fold average of the equation of state between the end of inflation and the end of reheating and $g_{\textrm{re}}$ is the number of effective bosonic degrees of freedom at the end of reheating. Using the continuity equation, one can write the number of e-folds at the end of reheating, $N_{\textrm{re}}$ as
 \bea
 N_{\textrm{\textrm{\re}}}=\frac{-1}{3(1+\omega_{\textrm{\intt}})}\ln \left(\frac{\rho_{\textrm{\textrm{\re}}}}{\rho_{\textrm{\eend}}}\right)~.\label{1611}
 \eea
Putting \eqref{1611} in \eqref{16}, after some calculations and simplifications one can obtain $N_{\textrm{re}}$ as a function of  model dependent parameters $N_k$, $V_k$ and $\rho_{\textrm{end}}$
 \bea
 N_{\re}=\frac{4}{1-3\omega_{\intt}}\displaystyle \left[66.9- N_k - \ln \left(\frac{k}{a_0 H_0}\right)+\frac{1}{4}\ln \left(\frac{V_k^2}{M_{\textrm{Pl}}^4 \rho_{\eend}}\right)
-\frac{1}{12}\ln g_{\re}\right]~.\label{16111}
\eea
On the other hand, $\rho_{\re}$ is related to the reheating temperature, $T_{\re}$ through $\rho_{\textrm{re}}=(\pi^2/30) g_{\textrm{re}}T_{\textrm{re}}^4$
and $\rho_{\eend}$ depends on the potential energy at the end of inflation, $V_{\eend}$ via $\rho_{\textrm{end}}=(3/2)V_{\textrm{end}}$, obtained by setting $\omega_{\eend}=-1/3$. Inserting $\rho_{\re}$ and $\rho_{\eend}$ into \eqref{1611} and inverting it, one obtains $T_{\re}$ as
 \bea
 T_{\textrm{re}}=\left[\frac{45}{\pi^2 g_{\textrm{re}}}V_{end}\right]^{1/4} e^{-3(1+\omega_{\textrm{int}})N_{\textrm{re}}/4}.\label{17}
\eea

\begin{figure}[h]
\epsfig{figure=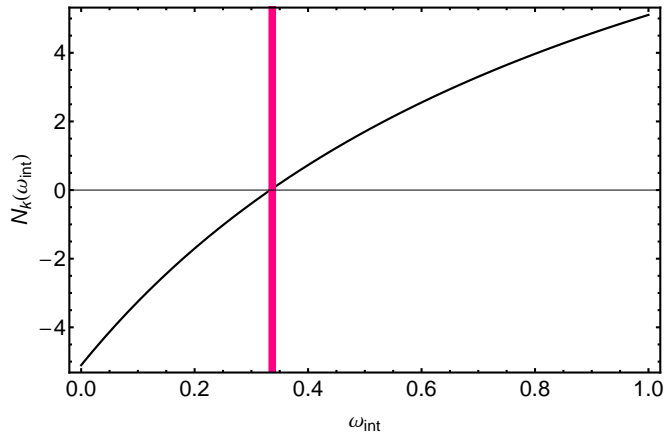,width=8.75cm}\hspace{2mm}
\caption{The number of e-folds, $N_k$ as a function of the reheating equation of state parameter $\omega_{\intt}$ for T-models. The vertical pink line show  $\omega_{\intt}=1/3$.}
 \label{fig4}
\end{figure}

Reheating should occur after the inflation and before BBN, i.e. $10^{-2}\:\textrm{GeV} \lesssim T_{\re} \lesssim 10^{16}\:\textrm{GeV}$. In addition, the allowable range for $\omega_{\intt}$ is between $-1/3$, from the end of inflation condition and $1$, to satisfy the positivity energy condition. Of course, it is difficult to consider $\omega_{\intt}>1/3$ since it requiers an unnatural inflaton field of the order of higher than $\phi^6$ \cite{Dai:2014jja}. It is well known that $\omega_{\intt}$ of the reheating phase for large field models is given by \cite{Turner,Kofman:1997yn,Martin:2003bt,Martin:2010}
\beq
\omega_{\intt}=\dfrac{p-2}{p+2}~, \label{1711}
\eeq
where $p$ is the power of the inflaton field in the corresponding potential. During the reheating era, one can check that the inflaton fields in T- and E-models with the powers $n=1/2$, $n=1$ and $n=2$ around their minimum behave as the large fields $\varphi$ ($p=1$), $\varphi^2$ ($p=2$) and $\varphi^4$ ($p=4$), respectively. Therefore, considering \eqref{1711} the apprpopriate values for the reheating equation of state parameter corresponding to T- and E-models with the noted powers are $\omega_{int}=-1/3$, $\omega_{\intt}=0$ (canonical reheating) and  $\omega_{int}=1/3$, respectively. But as \eqref{16111} and \eqref{17} are not well defined for $\omega_{\intt}=1/3$, we could just choose $-1/3 \leq\omega_{\intt}<1/3$ for our investigation. Of course, for $\omega_{\intt}=1/3$ it is clear that \eqref{16} is reduced to
\bea
 N_k=66.9- \ln \left(\frac{k}{a_0 H_0}\right)+\frac{1}{4}\ln \left(\frac{V_k^2}{M_{\textrm{Pl}}^4}\rho_{\textrm{end}}\right)
-\frac{1}{12}\ln g_{\textrm{re}} ~,\label{17111}
\eea
and at least one can obtain a prediction for $N_{\textrm{k}}$ and $n_{\textrm{s}}$ of a specific model.
\\
\indent
Before going through reheating analysis of T- and E-models, we want to study numerically the variation of the value of $N_k$ as a function of $\omega_{int}$ for the same reheating temperature, which can be taken for definiteness, $10^{-10} M_p$. For this purpose, we consider the $\omega_{\intt}$-dependent part of $N_{\textrm{k}}$, i.e.  the term before the last in \eqref{16} as
\bea
 N_k(\omega_{\intt})=\frac{1-3\omega_{\textrm{int}}}{12(1+\omega_{\textrm{int}})}\ln\left(\frac{\rho_{\textrm{re}}}{\rho_{\textrm{end}}}\right)~,\label{ab}
\eea
and plot it against $\omega_{\intt}$ for T-models. Fig. \ref{fig4} gives a big difference by $\Delta N_k\simeq 5$ when $\omega_{\intt}$ grows from 0 to 1/3. For example, what was $N_k = 54$ becomes $N_k = 59$. Now, one can use the simplest approximation $n_s = 1-2/N_k$ and find that e.g. for $N_k = 54$ this $\Delta N_k$ results in the change from $n_s = 0.963$ (in this approximation) to $0.9663$. So it is a big shift by $0.0033$ which moves the T-model to the safe $1\sigma$ confidence region completely. Therefore, we coclude that $\alpha\,$-attractors with $n=2$ and $\omega_{\intt}=1/3$ are more confident than ones with $n=1$ and $\omega_{\intt}=0$.

In this work, we set $g_{\re} = 106.75$ and use the pivot point $k = 0.002\,\textrm{Mpc}^{-1}$ as used in section 6 of the Planck 2015 inflation paper. So far, we have derived the reheating parameters $N_{\re}$ and $T_{\re}$ in terms of the model dependent quantities of inflation $N_k$, $V_{k}$ and $V_{\textrm{end}}$. In the following subsections, we will study the behavior of $N_{\re}(n_\s)$ and $T_{\re}(n_\s)$ to obtain reheating constraint on $\alpha$ parameter.

\subsection{T-models }

\begin{figure}[h]
\epsfig{figure=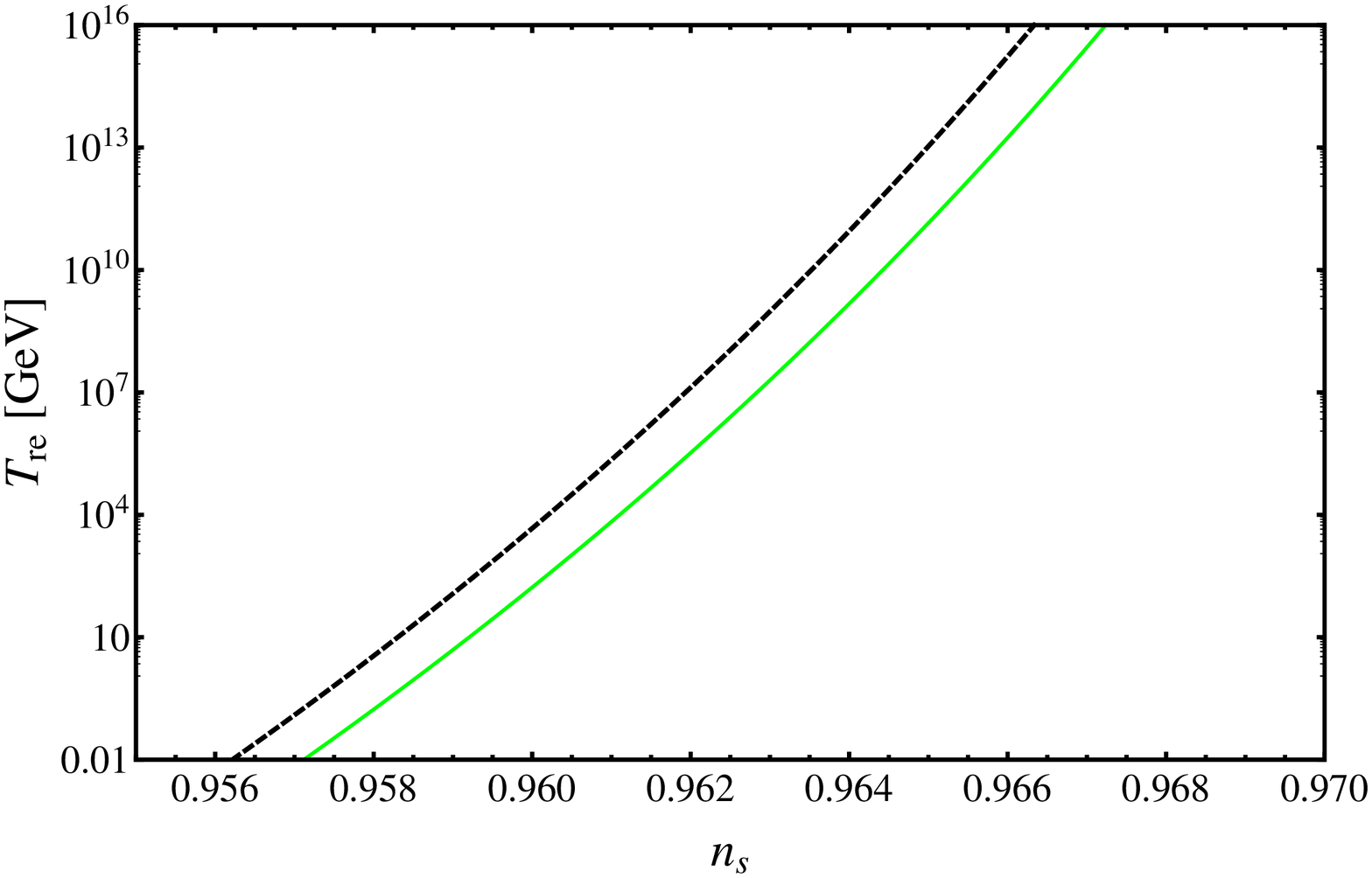,width=8.75cm}\hspace{2mm}
\epsfig{figure=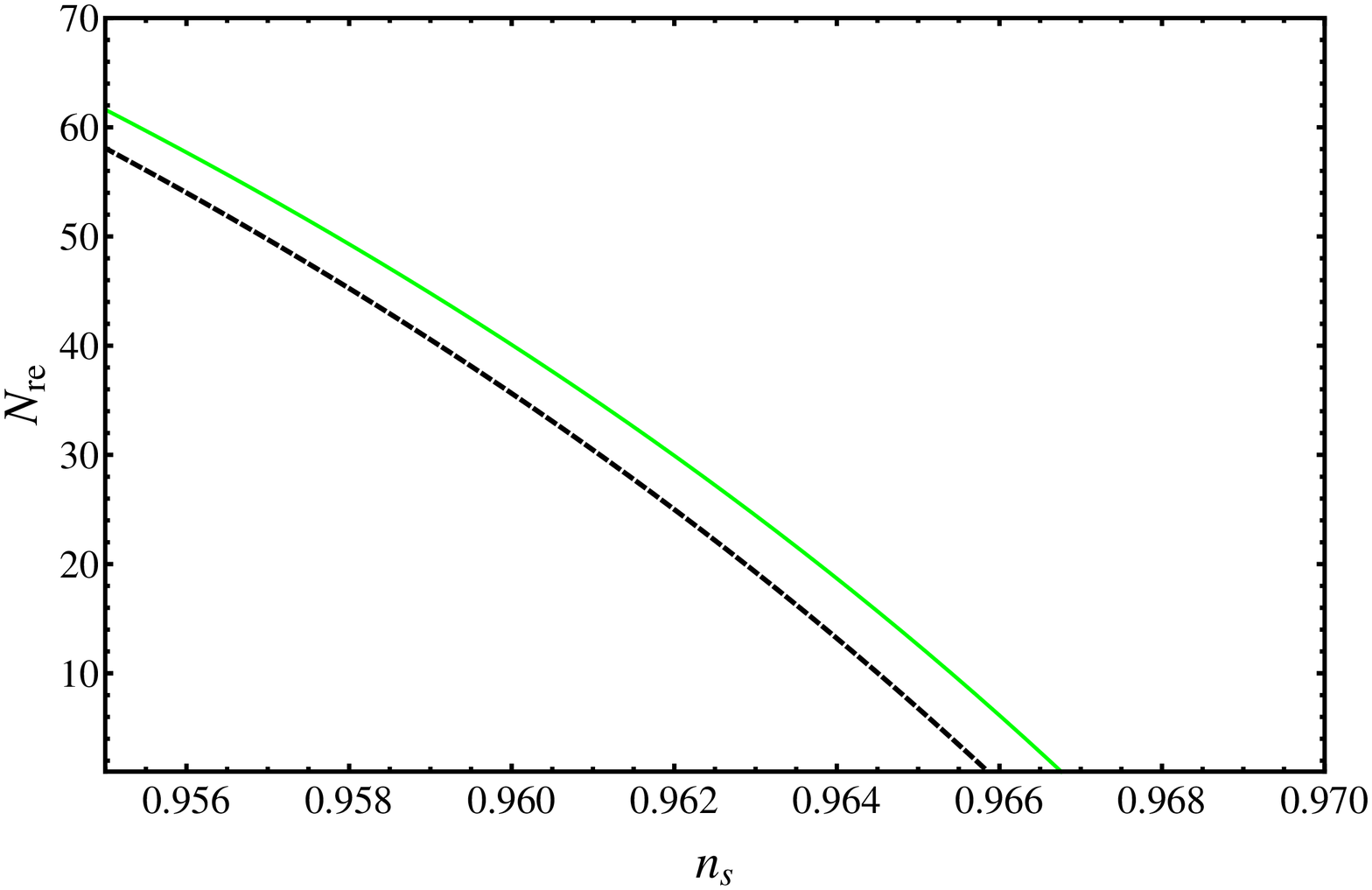,width=8.75cm}\hspace{2mm}
\caption{The reheating temperature, $T_{\re}$ (left panel) and the reheating number of e-folds, $N_{\re}$ (right panel) as functions of the spectral index, $n_\s$ for T-models. The green solid lines show $T_{\re}$ and $N_{\re}$ with respect to $n_\s$ with numerical correction term $\delta$ while the black dashed ones are $T_{\re}$ and $N_{\re}$ against $n_\s$ due to the slow-roll approximation i.e. without $\delta$.}
 \label{fig5}
\end{figure}

Here, we calculate the model dependent parameters in \eqref{16111} and \eqref{17} for T-model potential \eqref{Tmodelpotential} with $n=1, 2$. First, by inverting \eqref{184}, one can find $N_k$ in terms of $n_\s$ as
 \beq
 N_k=\frac{2}{1+\delta - n_\s}~. \label{22}
 \eeq
Then, we find the value of the inflaton field at the time of horizon crossing, $\varphi_k$,  by integrating \eqref{18} and  inverting its result
  \beq
  \varphi_k =\sqrt{\frac{3\alpha}{2}} M_{\textrm{Pl}}\cosh^{-1}\left(\frac{4 n N_k}{3 \alpha}+ \cosh(\sqrt{\frac{2}{3\alpha}}\,\varphi_{\textrm{\eend}}/M_{\textrm{Pl}})\right)~. \label{23}
  \eeq
On the other hand, using \eqref{185} we find $\varphi_{\eend}$ for the allowable range of $\alpha$ given in the previous section
\bea
&&\varphi_{\eend}=0.84~M_\textrm{Pl}~,\:\:\:\:\:\:\:   \textrm{for}~\:\:\:\:\:\:\:   \alpha=1~,\:n=1~,\\
&&\varphi_{\eend}=1.52~M_\textrm{Pl}~,\:\:\:\:\:\:\:   \textrm{for}~\:\:\:\:\:\:\:   \alpha=1~,\:n=2~,\\
&&\varphi_{\eend}=2.10~M_\textrm{Pl}~,\:\:\:\:\:\:\:    \textrm{for}~\:\:\:\:\:\:\:   \alpha=7~,\:n=2~,\\
&&\varphi_{\eend}=1.01~M_\textrm{Pl}~,\:\:\:\:\:\:\:    \textrm{for}~\:\:\:\:\:\:\:   \alpha=25~,\:n=1~.\label{24}\
\eea
Since we do not know the exact value of $\lambda_{n}$ in \eqref{Tmodelpotential}, it is reasonable to compute $V_{\textrm{end}}$ via $V_k$
  \beq
 V_{\textrm{\eend}}=V_{k}\,\left(\frac{\tanh(\varphi_{\textrm{\eend}}/(\sqrt{6\alpha}M_{\textrm{Pl}}))}{\tanh(\varphi_{k}/(\sqrt{6\alpha}M_{\textrm{Pl}}))}\right)^{2n}~,\label{26}
 \eeq
 where $V_{k}$ is the energy scale of inflation \cite{Ade:2015lrj}
\beq
V_{k}\approx (1.88\times 10^{16}\:\textrm{GeV})^{4}\,\frac{r}{0.10}~.\label{27}
\eeq
Finally, we substitute expressions (24-31) into \eqref{16111} and \eqref{17} to obtain $N_{\textrm{re}}$ and $T_{\textrm{re}}$. Before a full investigation, we apply the results for the simplest case of $n=1$ and $\alpha=1$ to examine the effects of correction terms in expression \eqref{184} on $N_{\re}$ and $T_{\re}$ comparing with slow-roll predictions. In Fig. \ref{fig5} we have shown the dependence of $T_{\textrm{re}}$ and $N_{\textrm{re}}$ on $n_\s$, comparing analytical approximations
with exact numerical calculations. Our results indicate that the numerical correction term $\delta$ leads to an excess for $N_{\textrm{re}}$ and a significant downfall in $T_{\textrm{re}}$. However, the effect of the numerical correction term $\lambda$ on $T_{\textrm{re}}$ and $N_{\textrm{re}}$ is negligible.

\begin{figure}[h]
\epsfig{figure=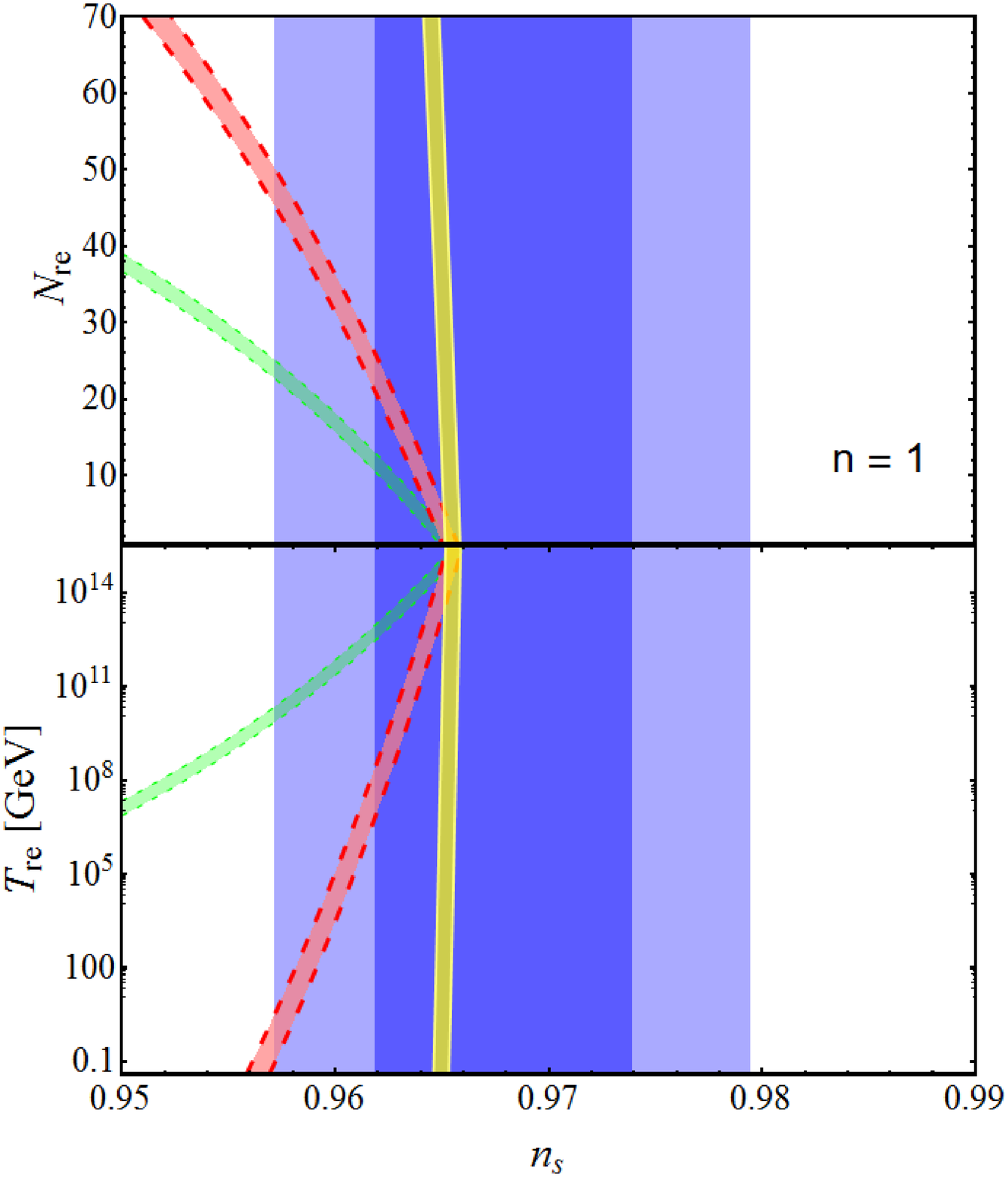,width=8.75cm}\hspace{2mm}
\epsfig{figure=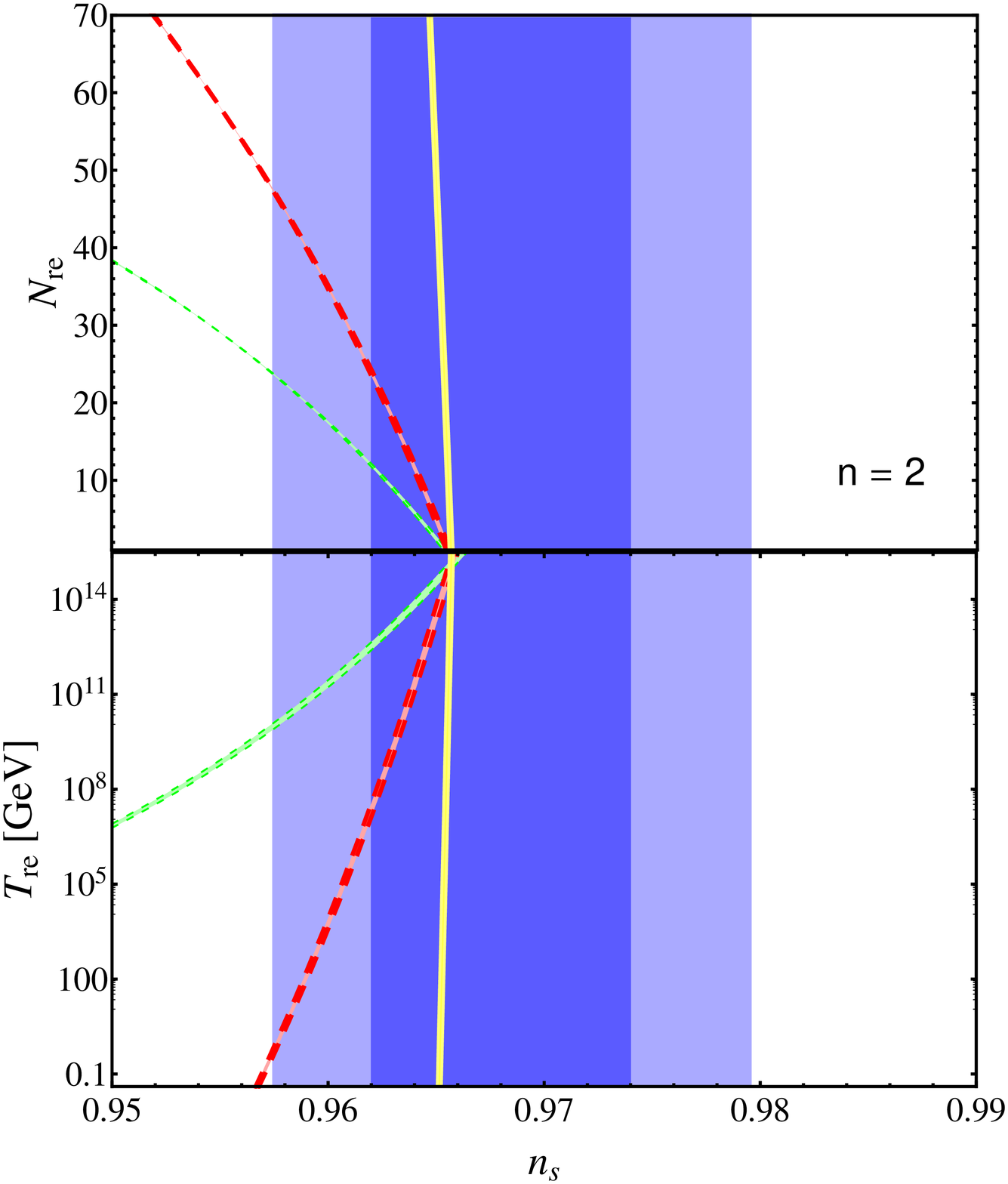,width=8.75cm}\hspace{2mm}
\caption{T-models: Plots of $T_{\textrm{re}}$ and $N_{re}$ as functions of $n_s$. In each plot, the green dotted, the red dashed and the yellow solid curves correspond to $\omega_{\textrm{re}}=-1/3$, $\omega_{\textrm{re}}=0$ and  $\omega_{\textrm{re}}=0.3$ respectively. Moving from left to right, the width of these curves correspond range of variation of the free parameters of model as $1\leq \alpha \leq25$ for $n=1$ and $1\leq \alpha \leq7$ for $n=2$. The dark blue band and the light blue one represent the 1$\sigma$ and 2$\sigma$ regions of Planck \cite{Planck:2015xua} and BICEP2/Keck \cite{Array:2015xqh} results, respectively.}
 \label{fig6}
\end{figure}

In the following, we consider the correction term in \eqref{22} and plot $T_{\textrm{re}}$ and $N_{\textrm{re}}$ for T-models with $n=1, 2$ and $\omega_{\textrm{int}}=-1/3,\ 0,\ 0.3$. As the Fig. \ref{fig6} shows, variation of $T_{\textrm{re}}(n_{\s})$ and $N_{\textrm{re}}(n_{\s})$ of T-models with $n=1, 2$ has a weak dependence on $\alpha$. \\
\indent
To have a better interpretation of Fig. \ref{fig6}, let us first expand T-model potentials around their minimum. It is mentioned before that just after end of inflation, i.e. the early stage of reheating era, these potentials with $n=1$ and $n=2$ are proportional to $\varphi^2$ $(p=2)$ and $\varphi^4$ $(p=4)$, respectively. Therefore, \eqref{1711} gives their corresponding reheating equation of state parameter as $\omega_{\textrm{int}}=0$ for models with $n=1$ and $\omega_{\textrm{int}}=1/3$ for ones with $n=2$ almost immediately after the end of inflation. From the left panel of Fig. \ref{fig6}, remaining within the $1\sigma$ confidence intervals requires $T_{\re}\gtrsim10^{7}$ GeV and $N_{\re}\lesssim25$ for models with $n=1$ and $\omega_{\textrm{int}}= 0$. For $\omega_{\textrm{int}}=1/3$ which corresponds to a vertical line passing through the intersection point of curves, our plots put no special constraint on reheating temperature. However, considering the $1\sigma$ confidence region, Fig. \ref{fig6} shows us that for bigger $\omega_{\textrm{int}}$, lower reheating temperatures are accesible. Therefore, we conclude that low reheating temperatures probable for models with $n=2$ and $\omega_{\textrm{int}}=1/3$.
\\
\indent
It is useful to compare the above results of T-models with reheating results of Higgs model. The Higgs potential which is extremely flat during inflation (similar to T-models) has a quadratic form just after end of inflation. In other words, it is quadratic until the amplitude of the oscillations drops down by 5 orders of magnitude \cite{Ferrara:2010in}. So during the inflaton oscillations, one has $\omega_{\textrm{int}}=0$ (as for quadratic potential) for Higgs potential, until one of the two things happens; reheating ends or the field becomes much smaller to obtain a quartic form. Thus in Higgs scenario one may have $\omega_{\textrm{int}}=1/3$ even if reheating occurs much later but typically, reheating happens earlier. Considering $T_{\textrm{re}}^{\textrm{Higgs}}\sim6\times10^{13}$ GeV \citep{Bezrukov:2008ut} and $\omega_{\textrm{int}}=0$ (canonical reheating), one can obtain \cite{Bezrukov:2011gp}
\beq
N_k=57.66~,\:\:\:\:\:\:\:\:\:\:\:\:\:\: n_s=0.967~,\label{27.5}
\eeq
If one supposes longer reheating, T-model potential with $n=1$ which has a quadratic form at the begining of the reheating becomes quartic with $\omega_{\textrm{int}}=1/3$ later. So it has a nearly similar behavior to Higgs potential during reheating and predicts a high reheating temperature as discussed before. Meanwhile, for T-models with $n=2$, the potential becomes quartic immediately after inflation with $\omega_{\textrm{int}}=1/3$. Therefore, T-models with $n=2$ predict low reheating temperature. On the other hand, based on theories in which gravitinos are produced during reheating, a delayed reheating leads to a very high temperature $T_{\re}\sim10^{14}$ GeV and overproduction of gravitinos and consequently overpopulation of the Universe with dark matter particles \cite{Bolz:2000fu,Steffen:2006hw,Cyburt:2009pg}. Furthermore, using \eqref{17111} and \eqref{184} we obtain
\beq
N_k=59.2~,\:\:\:\:\:\:\:\:\:\:\:\:\:\: n_s=0.9657~.\label{27.6}
\eeq
for T-models with $n=2$. This $n_s$ matches with intersection point of the curves which shows instantaneous reheating. Therefore, T-models with $n=2$ are good models which should help a lot more than Higgs model; their $n_s$ matches Planck data well (as discuused for Fig. \ref{fig4}) and they also give a low reheating temperature without conflict with possible existence of light gravitino.
\\
\indent
Of course, in order to find a spectral index efficiently close to the Planck best fit $n_\s=0.968$, T-models would make necessary $\omega_{\textrm{int}}>1/3$. 

\subsection{E-models}

We now turn to the reheating constraint on E-models.  To determine $N_{\textrm{re}}$ and $T_{\textrm{re}}$ in terms of spectral index, we first calculate $\varphi_k$ by substituting the Starobinsky-like potential \eqref{5} and its derivative in expression \eqref{18}
   \beq
  \varphi_k=\sqrt{\frac{3\alpha}{2}} M_{\textrm{Pl}} \ln \left(\frac{4n N_k }{3\alpha}\right)~. \label{29}
  \eeq
One can also obtain $\varphi_{\textrm{\eend}}$ by employing \eqref{185}
\bea
&&\varphi_{\eend}=0.614 M_\textrm{Pl}~,\:\:\:\:\:\:\:   \textrm{for}~\:\:\:\:\:\:\:   \alpha=1,\\
&&\varphi_{\eend}=0.778 M_\textrm{Pl}~,\:\:\:\:\:\:\:    \textrm{for}~\:\:\:\:\:\:\:   \alpha=5~.\label{31}\
\eea

\begin{figure}[h]
\epsfig{figure=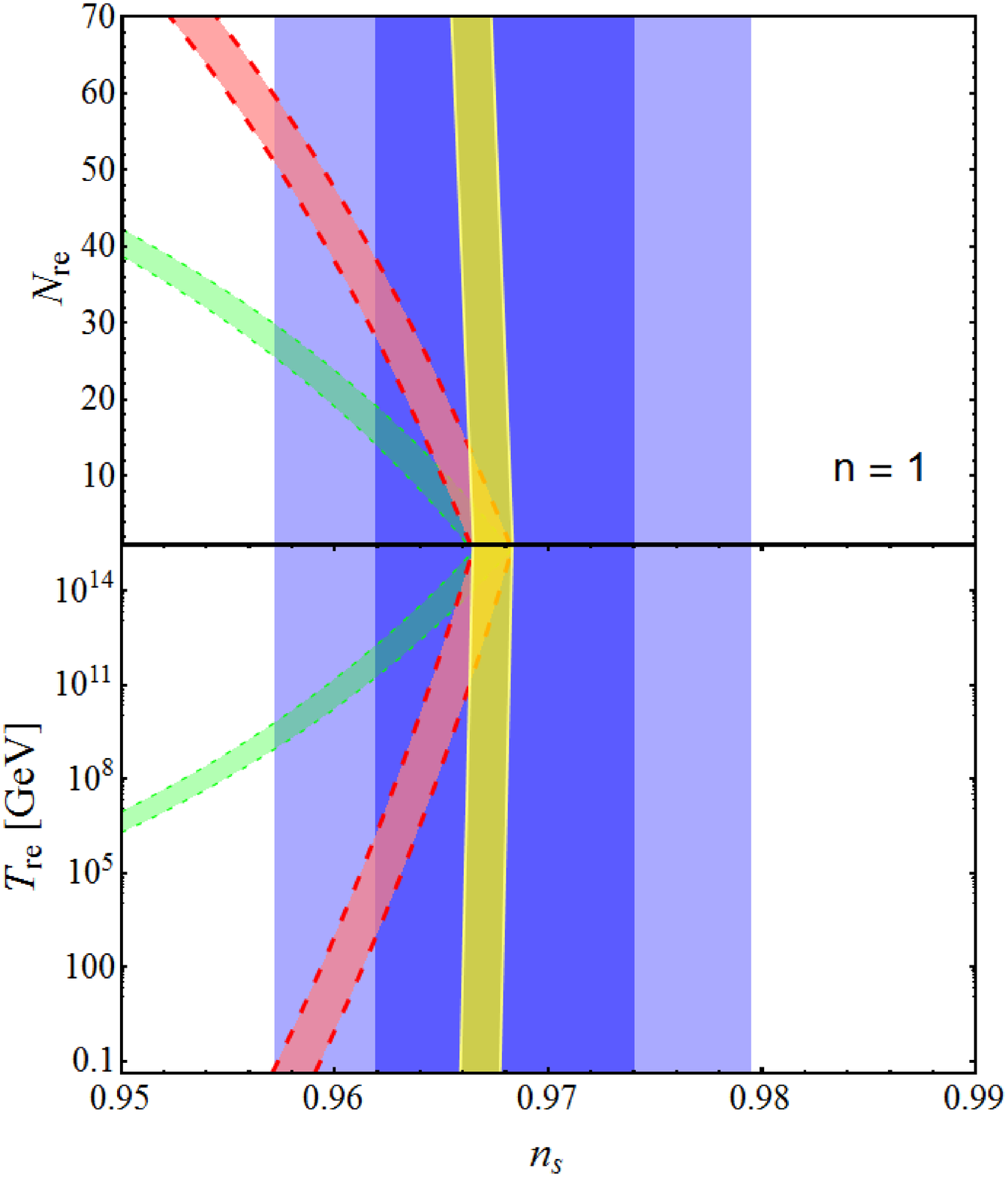,width=8.75cm}\hspace{2mm}
\caption{E-models: Plots of $T_{\textrm{re}}$ and $N_{re}$ as functions of $n_s$. The green dotted, the red dashed and the yellow solid curves correspond to $\omega_{\textrm{re}}=-1/3$, $\omega_{\textrm{re}}=0$ and  $\omega_{\textrm{re}}=0.3$ respectively. Moving from left to right, the width of these curves corresponds to the range of variation of the free parameters of model as $1\leq \alpha \leq5$ for $n=1$. The dark blue band and the light blue one represent the 1$\sigma$ and 2$\sigma$ regions of Planck \cite{Planck:2015xua} and BICEP2/Keck \cite{Array:2015xqh} results, respectively.}
 \label{fig7}
\end{figure}

Now using the expressions derived for $\varphi_k$ and $\varphi_{\textrm{end}}$, one can get $V_{\textrm{end}}$ as
  \beq
 V_{\textrm{end}}=3M^{2}_{\textrm{Pl}}H^{2}_{k}\,\left(\frac{1- e^{-\sqrt{\frac{2}{3\alpha}}\varphi_{\textrm{end}}/M_{\textrm{Pl}}}}{1- e^{-\sqrt{\frac{2}{3\alpha}}\varphi_{\textrm{end}}/M_{\textrm{Pl}}}}\right)^{2n}~.\label{32}
 \eeq
Inserting the derived expressions above and \eqref{22} into \eqref{16111} and \eqref{17}, one can calculate and plot $T_{\textrm{re}}$ and $N_{\textrm{re}}$ for E-models as a function of $n_s$.\\
\indent
We plot $T_{\textrm{re}}$ and $N_{\textrm{re}}$ for E-models with $n=1$ and $1\leq \alpha \leq5$ in Fig. \ref{fig7}. As one can see, variation of $T_{\textrm{re}}$ and $N_{\textrm{re}}$ is more dependent on parameter $\alpha$ compared to T-models. In the case of $n=1$ and $\alpha=1$, i.e. the Starobinsky model, using $T_{\re}^{R^2}\sim3\times10^{9}$ GeV \cite{Gorbunov:2010bn} and supposing post-inflationary matter dominated stage (i.e. $\omega=0$) we obtain
 \beq
  N_k^{R^2}=54.6~,\:\:\:\:\:\:\:\:\:\:\:\:\:\: n_\s^{R^2}=0.964 \label{33}
  \eeq
which are consistent with the analytical slow-roll results have been reported in \cite{Bezrukov:2011gp}.\\
\indent
In Fig. \ref{fig7}, the intersection line of the curves, where the equation of state parameter is irrelevant, shows instantaneous reheating which leads to zero number of e-folds and maximum temperature for the reheating. For $\alpha=5$, the right end point of this line corresponds to the Planck 2015 best fit $n_s= 0.968$. Therefore, $\alpha> 5$ is required for having a longer reheating with lower temperature. Considering small reheating temperature $T_{\textrm{re}}\lesssim 10^7$ GeV, which is suggested by those who believe light gravitino \cite{Cyburt:2002uv}, Fig. \ref{fig7} suggests E-models with $\alpha> 5$. However, if one suppose that gravitino are superheavy and there are no special constraints on reheating temperature \cite{Bolz:2000fu,Steffen:2006hw,Cyburt:2009pg}, then E-models with different values of $\alpha$ are possible.\\
\indent

\section{Conclusion }

It has been found recently that the $\alpha$-attractors provide a large class of inflationary models which is in good agreement with recent CMB data. On the other hand, it is investigated that reheating analysis can help to break the degeneracy of similar predictions of inflationary models. Using the recent CMB data of Planck 2015 and BICEP2/Keck 2015 and also considering the constraints that reheating puts on inflationary models, we studied two classes of $\alpha$-attractors, the so-called T- and E-models.\\
\indent
At first, we applied the exact numerical solutions of background equations for the $\alpha$-attractors to remove the imprecision of the slow-roll approximation for the next calculations. Comparing the results of this approach with the new constraints on cosmological observables, we confined the parameter $\alpha$ for both T- and E-models. We concluded that the upper bound $r_{0.05}< 0.07$ (95\% CL) \cite{Array:2015xqh} requires $\alpha\leq25$ and $\leq7$ for T-models with $n=1$ and 2, respectively. This investigation also indicated that E-models are very safe for $\alpha>5$ as compared to the $1\sigma$ region of $n_s-r$ plot of Planck 2015. As a comparison, we understood that for a specific range of $\alpha$, $r$ varies slower for E-models  compared to T-models.\\
\indent
Then, we presented a definite relationship between reheating parameters $T_{\textrm{re}}$, $N_{\textrm{re}}$ and $n_s$ for both models but we did not confine our study to canonical reheating. Our analysis indicated that the exact numerical approach implies a reheating temperature smaller by $O(10)$ and a reheating number of e-folds bigger by $O(1)$ than those are obtained from slow-roll approximation.\\
\indent

 For the canonical reheating, T-models with $n=1$ suggest $T_{\re}\gtrsim10^{7}$ GeV and $N_{\re}\lesssim25$ considering $1\sigma$ confidence region. However, as bigger $\omega_{\textrm{int}}$ could give bigger $n_s$ completely within $1\sigma$ confidence region (Fig. \ref{fig4}) and lower reheating temperatures (Fig. \ref{fig6}) without conflict with possible existence of light gravitino, we conclude that T-models with $n=2$ and $\omega_{\textrm{int}}=1/3$ looks more helpful than Higgs model. In the case of E-models, $\alpha> 5$ is required to respect the belief of light gravitino which leads to low reheating temperature. However, superheavy gravitino which predicts high reheating temperature puts no special constraints on E-models. For E-models with $n=1$ and $\alpha=1$, i.e. the Starobinsky model, our results are compatible with the analytical results of \cite{Bezrukov:2011gp}.\\
 \indent
None of these conclusions are fully reliable, because one standard deviation is not really a real constraint. However, one can look at these conditions and say that further investigation of $n_s$ and of consequences of reheating are important since they may strengthen or relax some of these conclusions.

\section*{\small Acknowledgement}

The authors would like to thank A. Linde specifically for his helpful discussions which lead to improvement of this work. M. Eshaghi thanks V. Domcke, P. Creminelli and C. Baccigalupi for various comments and useful discussions. M. Eshaghi acknowledges C. Baccigalupi and the organizers of Astrophysics sector of the International School for Advanced Studies, SISSA for their hospitality during the completion of this work.


\end{document}